# An Accurate Bilinear Cavern Model for Compressed Air Energy Storage

Junpeng Zhan[a,b], Osama Aslam Ansari[a], Weijia Liu[a], C. Y. Chung[a]

[a]*Department of Electrical and Computer Engineering, University of Saskatchewan, Saskatoon, SK S7N 5A9, Canada*

[b]*Sustainable Energy Technologies Department, Brookhaven National Laboratory, Upton, NY 11973, U.S.A.*

*Abstract*—**Compressed air energy storage is suitable for large-scale electrical energy storage, which is important for integrating renewable energy sources into electric power systems. A typical compressed air energy storage plant consists of compressors, expanders, caverns, and a motor/generator set. Current cavern models used for compressed air energy storage are either accurate but highly nonlinear or linear but inaccurate. The application of highly nonlinear cavern models in power system optimization problems renders them computationally challenging to solve. In this regard, an accurate bilinear cavern model for compressed air energy storage is proposed in this paper. The charging and discharging processes in a cavern are divided into several real/virtual states. The first law of thermodynamics and ideal gas law are then utilized to derive a cavern model, i.e., a model for the variation of temperature and pressure in these processes. Thereafter, the heat transfer between the air in the cavern and the cavern wall is considered and integrated into the cavern model. By subsequently eliminating several negligible terms, the cavern model reduces to a bilinear model. The accuracy of the bilinear cavern model is verified via comparison with both an accurate nonlinear model and two sets of field-measured data. The bilinear cavern model can be easily linearized and is then suitable for integration into optimization problems considering compressed air energy storage. This is verified via comparatively solving a self-scheduling problem of compressed air energy storage using different cavern models.**

## Nomenclature

**Parameters**

| | |
|---|---|
| a,ht | Both adiabatic process and heat transfer are considered (superscript) |
| $a_i$ | Parameters, $i = 1,2,3,\cdots,13$, given in the Appendix |
| $c_{Ain}$ / $c_{Aout}$ | Coefficients used to model the relationship between charging/discharging power and mass flow rate in/out |
| $c_i$ | Parameters, $i = 0,1$, representing the left-hand side of (3) and (10), respectively, given in the Appendix |
| $c_j$ | Parameters, $j = 2,3,\cdots,32$, given in the Appendix |
| $c_v$ | Constant volume specific heat (J/(kg K)) |
| ht | Heat transfer (superscript) |
| $h_c$ | Heat transfer coefficient (W/(m² K)) |
| $k$ | A constant equal to 1.4 |
| $l_x$ | Parameters used in (35), (36), (47), and (48), $x = 1,2,3,4$ |
| $m_o$ | Mass of air in virtual container 2 as shown in Fig. 3 (kg) |
| $m_{in}$ | Mass of air charged into the cavern (kg) |
| $n_t$ | Number of time steps |
| $p_{in}$ | Pressure of the air charged into the cavern (bar) |
| $p_{inj}$ | Pressures in virtual states as shown in Fig. 2, $j = 1,2$ (bar) |
| $p_{min}, p_{max}$ | Maximum and minimum pressures in a cavern for optimal operation of compressed air energy storage (CAES), respectively (bar) |
| $p_{si}$ | Pressure of the air in the cavern after the charging, discharging, and idle processes for $i = 2, 3,$ and 4, respectively (bar) |
| $t$ | Time step |
| $A_c$ | Surface area of the cavern wall (m²) |
| $C_{ch}^t, C_{dch}^t$ | Operational costs of charging and discharging power in CAES, respectively ($/MWh) |
| $P_{ch}^{max}, P_{ch}^{min}$ | Maximum and minimum charging power of CAES, respectively (MW) |
| $P_{dch}^{max}, P_{dch}^{min}$ | Maximum and minimum discharging power of CAES, respectively (MW) |
| $Q$ | Total internal energy (J) |
| $R$ | Specific air constant (J/(kg K)) |
| $T_{si}$ | Temperature of the air in the cavern after the charging, discharging, and idle processes for $i = 2, 3,$ and 4, respectively (K) |
| $T_{RW}$ | Temperature of the cavern wall (K) |
| $T_{in}$ | Temperature of the air charged into the cavern (K) |
| $T_{inj}$ | Temperatures in virtual states as shown in Fig. 2, $j = 1,2$ (K) |
| $V_{in1}, V_o$ | Volumes of virtual containers as shown in Figs. 2 and 3 (m³) |
| $V_s$ | Volume of a cavern (m³) |
| $W$ | Work (J) |
| $\rho_{av}$ | Average air density in a cavern (kg/m³) |
| $\tau_t$ | Electricity price ($/MWh) |
| $\Delta t$ | Time interval (s) |
| $\Delta U$ | Change in internal energy (J) |
| $\Omega_{T0}$ | $0,1,2,\cdots,(n_t - 1)$ |
| $\Omega_{T1}$ | $1,2,3,\cdots,n_t$ |

**Variables**

| | |
|---|---|
| $\dot{m}_{in}$ | Mass flow rate charged into a cavern (kg/s) |
| $\dot{m}_{out}$ | Mass flow rate discharged out of a cavern (kg/s) |
| $m_s$ | Mass of air in the cavern (kg) |

E-mail addresses: zhanjunpeng@gmail.com, oa.ansari@usask.ca, liuweijiamarcel@gmail.com, c.y.chung@usask.ca).

| | |
|---|---|
| $p_s$ | Pressure of air in the cavern (bar) |
| $p_{s,(xx)}^{(t+1)}$ | Pressure after a (xx) process, where (xx) can be 'ch', 'dch', and 'idl', which represent charging, discharging, and idle, respectively (bar) |
| $T_s$ | Temperature of air in the cavern (K) |
| $T_{s,(xx)}^{(t+1)}$ | Temperature after a (xx) process, where (xx) can be 'ch', 'dch', and 'idl', which represent charging, discharging, and idle, respectively (K) |
| $P_{ch}^t, P_{dch}^t$ | Charging and discharging power, respectively (MW) |
| $\alpha^t, \beta^t$ | Binary variable indicating the charging and discharging processes, respectively |

## I. INTRODUCTION

Energy storage technologies are critical to electric power systems, especially considering that the penetration of renewable generation is growing rapidly, e.g., the share of wind power in global electricity generation will increase from 4% in 2015 to 25-28% in 2050 [1]. Energy storage can provide various kinds of services [2], [3], e.g., electric energy time-shift, electric supply capacity, regulation, power reliability, etc. The current global installed energy storage capacity is approximately 141 GW, and an estimated 310 GW of additional capacity is needed in the United States, Europe, China, and India [4] to support the massive increase of renewable generation in the future. Two types of large-scale energy storage currently exist, i.e., pumped-hydro power storage and compressed air energy storage (CAES), which can be cost-effectively installed at the electricity grid scale. Hydropower technologies are mature, and many sites that were feasible for constructing pumped-storage hydropower already have hydropower plants in place [5]. Therefore, this paper focuses on CAES technology.

A CAES plant consists of compressors, turbines, a motor/generator set, and large repositories, e.g., underground salt caverns. Fig. 1 shows a schematic of the Huntorf CAES plant to depict the main components of CAES [6]. CAES operates in one of three modes, i.e., charging, discharging, and idle. When charging, the motor/generator set acts as a motor that is connected to the compressors via a clutch, as shown in the 'Charging circuit' part of Fig. 1. Low-cost electricity (e.g., off-peak electricity or wind power) is usually used by the motor to compress air to high pressure for storage in a large repository. When discharging, the motor/generator set acts as a generator that is connected to the turbines via another clutch, as shown in the 'Discharging circuit' part of Fig. 1. The compressed air is released from the repository and then combusted with fuel (e.g., natural gas) to drive the turbines. When idling, no air is injected into or released from the repository. Multiple stages of compressors (turbines) are typically used instead of a single stage to increase the efficiency, e.g., low- and high-pressure compressors (turbines) are used in the Huntorf CAES plant as shown in Fig. 1.

CAES is a high power and high energy storage technology and has relatively low capital, operational, and maintenance costs [3]. The power rating of a large-scale CAES plant can reach 300 or even 1000 MW, and the rated energy capacity can reach 1000 or even 2860 MWh [3]. Different definitions of efficiencies for CAES are discussed in [7], e.g., the round-trip efficiency of the CAES typically ranges from 66 to 82%. According to the Electric Power Research Institute (EPRI), about 75% of the continental U.S. has geologic sites suitable for CAES [7], [8]. Northern Europe is also replete with suitable salt deposits. For example, nearly 500 salt caverns are currently being used for natural gas storage. That is, it is feasible to install CAES in many different locations, which makes it a promising large-scale energy storage technology.

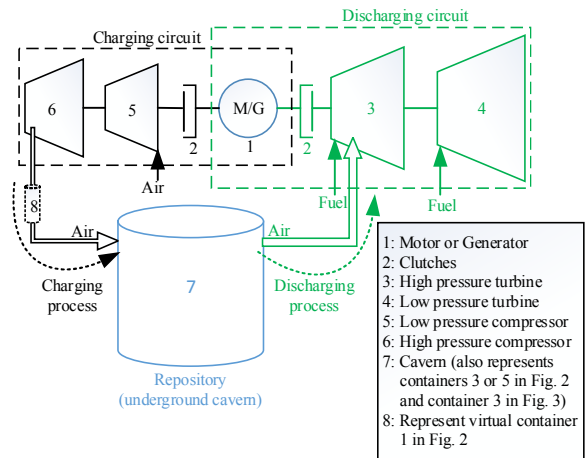

Fig. 1. Schematic of the Huntorf compressed air energy storage plant.

Two commercialized CAES plants are currently in operation. The world's first CAES plant was installed in Huntorf, Germany in 1978 [6]. The power ratings of its charging and discharging processes are 60 and 290 MW, respectively. The second commercialized CAES plant is the McIntosh plant in McIntosh, Alabama, U.S. [8], which became operational in 1991. The power rating of its charging process is 50 MW and it can produce an output of 110 MW electricity for up to 26 hours.

Due to several benefits offered by CAES, a number of CAES plants are being constructed for efficient utilization of renewable energy sources. For example, a feasibility study for a 160 MW CAES plant near the Saskatchewan-Alberta border in Canada [9] was finished in late 2018. This proposed CAES plant is expected to be combined with the interconnection between the Saskatchewan and Alberta power grids. From 2009 to 2013, Pacific Gas & Electric received US$50 million in funding for a demonstration project to validate the design, performance, and reliability of a 300 MW CAES plant in Kern County, California [10]. Several further examples are provided in [10], [11].

The two existing CAES plants in Huntorf and Alabama are diabatic CAES (D-CAES), and also referred to as the first generation CAES. Second generation CAES include adiabatic CAES (A-CAES) [12] [13], [14], [15] [16], [17], [18], isothermal CAES (I-CAES) [19], [20], and hybrid CAES [21], [22]. In the D-CAES, the heat generated in the compression process is wasted and an external heat source is required in the expansion process. In A-CAES, the heat generated in the compression process is captured, stored, and then used in the expansion process. I-CAES prevents the temperature from changing in the compression and expansion devices. Pacific Northwest National Laboratory (PNNL) and EPRI consider A-CAES and hybrid CAES to be the most suitable and promising CAES technologies [10]. For a complete review of CAES technologies, readers are referred to [10], which provides a comprehensive classification and comparison of different

                                                                    

CAES technologies as well as several challenging issues relevant for CAES research and development.

Air reservoir is a fundamental part of the CAES as it significantly affects the power and energy rating as well as the total construction and operation costs [11]. The reservoirs can be above-ground containers or underground caverns, including salt, hard rock, and porous rock layer caverns [11], [23]. The existing commercial CAES plants mentioned above both use underground salt caverns for air storage. The proposed cavern model is general and can be applied to the different kinds of reservoirs mentioned above.

Some research has been conducted to model compressors and turbines [24], [25], [26]. However, the CAES cavern models available in the literature [11], [27], [28], [29], [30] are either too complicated or inaccurate and, as such, are not suitable for real applications. Therefore, this paper focuses on the cavern modeling of CAES, i.e., modeling the temperature and pressure variations of cavern air. Thermodynamic properties, such as variations of temperature and pressure in the caverns and the heat transfer between caverns and their surroundings, are important factors that affect the overall plant operation and performance [27], [28], [29]. Two kinds of cavern models for CAES are currently described in the literature.

The first kind is models that are accurate but highly nonlinear [27], [28], [29]. In [27], complex and simplified real gas models are developed for an adiabatic cavern for CAES, both of which adequately represent the thermodynamic properties of the air. Reference [28] developed an accurate dynamic simulation model for a CAES cavern that incorporates an accurate heat transfer model; the heat transfer is shown to play an important role in the thermodynamic behavior of the cavern and therefore the model in [28] can accurately simulate the actual cavern behavior. In [29], a simplified and unified analytical solution considering heat transfer is proposed for temperature and pressure variations in CAES caverns. This model is validated using data from the Huntorf plant trial tests and the results calculated from the models in [27] and [28], demonstrating that the solution in [29] is capable of adequately calculating the thermodynamic behavior of CAES caverns. Note that the models described in [27], [28], and [29] are accurate but highly nonlinear.

The second kind of cavern model assumes that the air temperature in the cavern is constant. This kind of model has been adopted in different power system operation problems, e.g., transmission congestion relief [31], bidding and offering strategy [32], and unit commitment [33]. An overview of different applications of technologies relevant to CAES is available in [34]. The constant-temperature model is linear but inaccurate, which can result in non-optimal or even infeasible solutions in practice.

Therefore, all existing CAES applications use simple but inaccurate cavern models. To achieve a balance between accuracy and complexity, this paper proposes an original and novel bilinear cavern model. The charging (discharging) process is divided into four (two) real/virtual states. The ideal gas law and the first law of thermodynamics [35] are used to model the pressure/temperature relationships between different charging/discharging states. The heat transfer between the cavern air and the cavern wall is also considered. A bilinear cavern model is then obtained by ignoring some very-small-value terms and using several linearization methods such as Newton's generalized binomial theorem and the first-order Taylor series approximation [36]. This original bilinear model is derived step-by-step in this paper. The advantages of the bilinear model over the existing model types mentioned above are twofold: 1) it is accurate, as will be verified in this paper, and 2) it is suitable for integration in power system optimization problems as explained in the following two paragraphs.

To consider CAES in an optimization problem, binary variables are required to indicate the charging, discharging, or idle status. That is, an optimization problem considering CAES is a mixed-integer programming problem. If the nonlinear cavern models given in [27], [28], [29] are used for CAES, an optimization problem considering CAES becomes a mixed-integer nonlinear programming (MINLP) problem. MINLP problems are usually difficult to be accurately solved in an acceptable computational time [37], [38], and therefore nonlinear cavern models such as those given in [27], [28], [29] are not suitable to be employed in an optimization problem. If the proposed bilinear cavern model is used for CAES, an optimization problem considering CAES becomes a mixed-integer bilinear programming (MIBLP) problem, which can easily be converted into a mixed-integer linear programming (MILP) problem [39], [40]. MILP problems are relatively easier to solve than MINLP problems [37], [41]. MILP problems can also be efficiently solved by several mature, off-the-shelf commercial solvers, e.g., CPLEX [42] and Gurobi [43]. Therefore, the bilinear cavern model is more suitable for integration into an optimization problem compared to the highly nonlinear cavern models given in [27], [28], [29]. The above discussion serves as the main motivation to propose the bilinear model.

A typical example of the previously mentioned optimization problem is as follows. A CAES plant can act as an independent participator in electricity markets. In this case, the self-scheduling of a CAES plant is an important optimization problem that needs to be solved to maximize its arbitrage revenue obtained from selling electricity to and buying electricity from the market [34], [44], [45]. In this paper, the day-ahead self-scheduling problem (SSP) of CAES is used to demonstrate the advantage of using the proposed bilinear model in an optimization problem compared to the highly nonlinear cavern models given in [29]. For the sake of comparison, both the proposed bilinear cavern model and the nonlinear cavern model in [29] are integrated into the SSP of CAES; both SSPs are MINLP problems and can be solved by an MINLP solver, e.g., BARON [46] (note that MIBLP is a kind of MINLP). Furthermore, the SSP of CAES using the proposed bilinear model is an MIBLP model; the MIBLP model is converted into an MILP model, which is then solved by an MILP solver, e.g., CPLEX.

In summary, the contributions of this paper include 1) proposing a novel accurate bilinear cavern model of CAES that is superior to existing linear but inaccurate or accurate but highly nonlinear cavern models from an optimization point of view, 2) verifying the accuracy of the proposed bilinear cavern model via comparison with both an accurate model and field-measured data, and 3) using the SSP of CAES to verify the superiority of using the bilinear cavern model compared to a highly nonlinear cavern model in the optimization problem.

The reminder of this paper is organized as follows. Section II details the deduction of the accurate bilinear cavern model. Section III presents the SSP of CAES and flowcharts for using the bilinear cavern model. Section IV verifies the effectiveness of the bilinear cavern model. Section V presents the conclusions drawn from the results.

## II. ACCURATE BILINEAR CAVERN MODEL FOR COMPRESSED AIR ENERGY STORAGE

The cavern of a CAES plant can operate in either constant-volume or constant-pressure mode. In this paper, constant-volume caverns are considered as they are used in existing CAES plants [10].

### A. Charging Process

In the charging process, a certain amount of air is injected by compressors into a cavern, as shown in Fig. 1. To facilitate the model deduction given in Sections II-A1 to II-A3, the charging process is divided into four states and the air is assumed to be stored in the (virtual) containers as shown in Fig. 2, where the five containers are indexed by the numbers in heptagons. The characteristics of the air in each container, including the pressure, volume, temperature, and mass, are shown in each container in Fig. 2. The values of the underlined notations, i.e., the pressure and temperature in containers 2, 4, and 5, are unknown while the values of the other notations are known from the actual CAES operation.

Containers 1, 2, and 4 are virtual while containers 3 and 5 represent the cavern before and after the air from the compressors is injected into it, respectively.

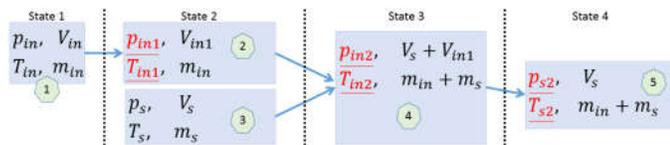

Fig. 2. Four states and five containers used for model deduction in the charging process.

The air coming out of the compressors is assumed to be stored in a virtual container, i.e., cylinder 8 in Fig. 1 and container 1 in Fig. 2. The conditions of container 1 represent the thermodynamic properties of air at the outlet of the compressors. First, let the air in virtual container 1 go into virtual container 2. The volume of container 2 is set such that the ratio of the volume of containers 2 to 3 is equal to the ratio of the mass of air in containers 2 to 3, i.e.,

$$V_{in1}/V_s = \dot{m}_{in}\Delta t/m_s \qquad (1)$$

where $V_{in1}$ and $V_s$ represent the volumes of containers 2 and 3, respectively; $m_s$ represents the mass of air in container 3; and $\dot{m}_{in}$ represents the flow rate of mass charged into the cavern, which is assumed to be constant during a period of time, $\Delta t$. The mass of air coming out of the compressor during that period of time, denoted as $m_{in}$, can be expressed as $m_{in} = \dot{m}_{in}\Delta t$. The mass of air injected into the cavern is assumed to be equal to $m_{in}$, i.e., there is no air leakage. $m_{in}$ is also the mass of air in both containers 1 and 2.

Then, let the air in both containers 2 and 3 go into container 4. The volume of (mass of air in) container 4 is set to the sum of the volumes of (masses of air in) containers 2 and 3, i.e., container 4 can be seen as a combination of containers 2 and 3. Note that the purpose of using virtual containers 2 and 4 is to let the work be zero during the process of merging the air in containers 2 and 3 into container 4.

Last, let the air in container 4 go into container 5, which is an adiabatic compression process as the mass of air does not change and the volume decreases. The rest of this subsection details the deduction of the model for the charging process.

#### 1) State 1 → State 2

The transfer of air from containers 1 to 2 is an adiabatic process and the mass of air does not change. According to the ideal gas law [35] for the air in container 2, one can obtain

$$p_{in1}V_{in1} = \dot{m}_{in}\Delta t RT_{in1}. \qquad (2)$$

According to the temperature-pressure relationship for an adiabatic process, the term $T^{\frac{k}{k-1}}/p$ is constant from containers 1 to 2 [47] and therefore one can obtain

$$\frac{(T_{in})^{\frac{k}{k-1}}}{p_{in}} = \frac{(T_{in1})^{\frac{k}{k-1}}}{p_{in1}}. \qquad (3)$$

Let $c_0$ represent the left-hand side of (3), i.e., $c_0 = T_{in}^{\frac{k}{k-1}}/p_{in}$. Note that the constant $c_0$ is known. $V_{in1}$ can be determined from (1), i.e., $V_{in1} = V_s\dot{m}_{in}\Delta t/m_s$. This leaves only two unknown variables in (2) and (3), i.e., $p_{in1}$ and $T_{in1}$. Therefore, $p_{in1}$ and $T_{in1}$ can be obtained from (2) and (3):

$$p_{in1} = (c_0)^{k-1}R^k m_s^k/V_s^k \qquad (4)$$

$$T_{in1} = (c_0 R m_s/V_s)^{k-1}. \qquad (5)$$

#### 2) State 2 → State 3

Now we consider the process of the air in both containers 2 and 3 going into container 4. The volume of container 4 is equal to the sum of the volumes of containers 2 and 3. In this process, the work is zero and the total internal energy [35] does not change. The change of the internal energy in containers 2 and 3 is $m_{in}c_v(T_{in2} - T_{in1})$ and $m_s c_v(T_{in2} - T_s)$, respectively. According to the first law of thermodynamics, i.e., $Q = \Delta U + W$, one can obtain

$$m_{in}c_v(T_{in2} - T_{in1}) + m_s c_v(T_{in2} - T_s) = 0. \qquad (6)$$

Note that $T_{in1}$ can be obtained from (5). $T_{in2}$ is then the only unknown variable in (6) and can be expressed as

$$T_{in2} = (m_{in}T_{in1} + m_sT_s)/(m_{in} + m_s). \qquad (7)$$

According to the ideal gas law for the air in container 4 in Fig. 2, one can obtain

$$p_{in2}(V_s + V_{in1}) = (m_{in} + m_s)RT_{in2}. \qquad (8)$$

By substituting (7) into (8), one can then obtain

$$p_{in2} = \frac{(m_{in}+m_s)RT_{in2}}{V_s+V_{in1}} = \frac{(m_{in}T_{in1}+m_sT_s)R}{V_s+V_{in1}}. \qquad (9)$$

Now $p_{in2}$ and $T_{in2}$ in container 4 have been obtained.

### 3) State 3 → State 4

From containers 4 to 5, the mass of air does not change and the volume reduces from $(V_s + V_{in1})$ to $V_s$. This is an adiabatic compression process. Thus, $T^{\frac{k}{k-1}}/p$ is constant from containers 4 to 5 [47], i.e.,

$$\frac{(T_{in2})^{\frac{k}{k-1}}}{p_{in2}} = \frac{(T_{s2})^{\frac{k}{k-1}}}{p_{s2}} \qquad (10)$$

According to the ideal gas law for the air in container 5 in Fig. 2, one has

$$p_{s2}V_s = (\dot{m}_{in}\Delta t + m_s)RT_{s2} \qquad (11)$$

Let $c_1$ represent the left-hand side of (10), i.e., $c_1 = (T_{in2})^{\frac{k}{k-1}}/p_{in2}$. Only two variables are unknown in (10) and (11), i.e., $p_{s2}$ and $T_{s2}$. Therefore, $p_{s2}$ and $T_{s2}$ can be obtained from (10) and (11):

$$p_{s2} = (\dot{m}_{in}\Delta t + m_s)^k R^k (c_1)^{k-1}/V_s^k \qquad (12)$$

$$T_{s2} = (c_1 p_{s2})^{1-\frac{1}{k}} . \qquad (13)$$

Now $p_{s2}$ and $T_{s2}$ for container 5 have been obtained. By substituting $c_1$ ($c_0$, (5), (7) and (9) are needed to calculate $c_1$) into (12) and (13), these two equations can be rewritten as:

$$p_{s2} = p_s\left(1 + \frac{\dot{m}_{in}\Delta t}{m_s}\right)^{k-1} + a_2(\dot{m}_{in}\Delta t + m_s)^{k-1}\dot{m}_{in}\Delta t \qquad (14)$$

$$T_{s2} = T_s\left(1 + \frac{\dot{m}_{in}\Delta t}{m_s}\right)^{k-2} + a_3(\dot{m}_{in}\Delta t + m_s)^{k-2}\dot{m}_{in}\Delta t \qquad (15)$$

where $a_2 = \frac{R^k T_{in}^k}{V_s^k p_{in}^{k-1}}$ and $a_3 = \frac{R^{k-1} T_{in}^k}{V_s^{k-1} p_{in}^k}$.

Equations (14) and (15) show that $p_{s2}$ and $T_{s2}$ are nonlinear functions of $\dot{m}_{in}$, which are linearized as follows. According to Newton's generalized binomial theorem [36], one has

$$(1+x)^r = \sum_{j=0}^{\infty}\binom{r}{j}x^j = 1 + rx + \frac{r(r-1)}{2!}x^2 + \cdots \qquad (16)$$

where $\binom{r}{j} = \frac{r(r-1)\cdots(r-j+1)}{j!}$, $r$ can be any real number, and $j$ is an integer. That is, $\left(1 + \frac{\dot{m}_{in}\Delta t}{m_s}\right)^{k-1}$ in (14) can be expressed as $\left(1 + (k-1)\frac{\dot{m}_{in}\Delta t}{m_s} + \frac{(k-1)(k-2)}{2!}\left(\frac{\dot{m}_{in}\Delta t}{m_s}\right)^2 + \cdots\right)$. Considering that $\dot{m}_{in}\Delta t$ is much smaller than $m_s$, the second and higher orders terms of $(\dot{m}_{in}\Delta t/m_s)$ can be ignored. Then, (14) can be reformed as

$$p_{s2} = p_s\left(1 + (k-1)\frac{\dot{m}_{in}\Delta t}{m_s}\right) + a_2(m_s)^{k-1}\dot{m}_{in}\Delta t \qquad (17)$$

Note that the second term in (14) is replaced by $a_2(m_s)^{k-1}\dot{m}_{in}\Delta t$, which has negligible error as $a_2$ is much smaller than $p_s$ (e.g., $p_s$ =46~66×$10^5$ Pascals and $a_2$ =1.04× $10^{-3}$ for the Huntorf CAES plant) and $\dot{m}_{in}\Delta t$ is much smaller than $m_s$. Similarly, (15) can be reformed as

$$T_{s2} = T_s\left(1 + (k-2)\frac{\dot{m}_{in}\Delta t}{m_s}\right) + a_3(m_s)^{k-2}\dot{m}_{in}\Delta t \qquad (18)$$

When (17) and (18) are used in a single-time-step optimization problem, $p_{s2}$ is a linear function of $\dot{m}_{in}$ in (17) and $T_{s2}$ is a linear function of $\dot{m}_{in}$ in (18) as $p_s$ and $T_s$ have a known initial status. When used in a multi-time-step optimization problem, (17) and (18) are bilinear equations as $p_s$ and $T_s$ become decision variables.

### B. Discharging Process

In the discharging process, the air is released from the cavern, as shown in Fig. 1. To facilitate the model deduction given in the rest of this subsection, the cavern before discharging is divided into two containers, i.e., containers 1 and 2, while the cavern after discharging (cylinder 7 in Fig. 1) is represented as container 3 in Fig. 3. The three containers in Fig. 3 are indexed by the numbers in heptagons. The characteristics of the air in each container, including the pressure, volume, temperature, and mass, are shown in each container in Fig. 3. The values of the underlined notations, i.e., the pressure and temperature in container 3, are not known while the values of the other notations are known.

The discharging process can be divided into two virtual steps. First, the air in container 2 is taken out of the cavern. Then, the air in container 1 expands to the whole cavern, i.e., air goes from containers 1 to 3. The deduction of the model for the second step is given in the rest of this subsection.

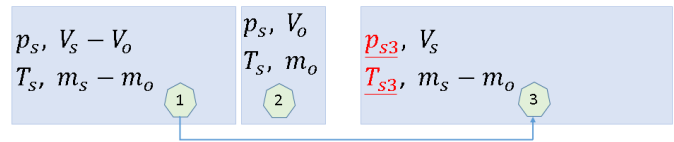

Fig. 3. Three containers used for model deduction in the discharging process.

The volume of container 2 is set such that the ratio of the volume of container 2 to that of container 1 is equal to the ratio of the mass of air in container 2 to that of container 1, i.e.,

$$V_o/(V_s - V_o) = m_o/(m_s - m_o). \qquad (19)$$

Note that the purpose of using virtual container 2 is to let the temperature and pressure of the air in container 1 be the same as the air in the cavern before discharging and to let air go from containers 1 to 3. This means that step 2 is an adiabatic expansion process.

Let $\dot{m}_{out}$ represent the flow rate of mass discharged from the cavern, which is assumed to be constant during a period of time, $\Delta t$. Then, the mass of air discharged from the cavern, denoted as $m_o$, during that period of time can be expressed as $m_o = \dot{m}_{out}\Delta t$.

The air expansion from containers 1 to 3 is an adiabatic process. Then, $T^{\frac{k}{k-1}}/p$ is constant from containers 1 to 3:

$$(T_s)^{\frac{k}{k-1}}/p_s = (T_{s3})^{\frac{k}{k-1}}/p_{s3}. \qquad (20)$$

According to the ideal gas law for the air in container 3 in Fig. 3, one has

$$p_{s3}V_s = (m_s - m_o)RT_{s3}. \qquad (21)$$

There are only two variables unknown in (20) and (21), i.e., $p_{s3}$ and $T_{s3}$. Therefore, $p_{s3}$ and $T_{s3}$ can be obtained from (20) and (21):

$$p_{s3} = (1 - \dot{m}_{out}\Delta t/m_s)^k p_s \qquad (22)$$

$$T_{s3} = (1 - \dot{m}_{out}\Delta t/m_s)^{k-1} T_s. \qquad (23)$$

Similar to the charging process, the above nonlinear equations are linearized. According to Newton's generalized binomial theorem, when $\dot{m}_{out}\Delta t$ is much smaller than $m_s$, (22) and (23) can be respectively reformulated as

$$p_{s3} = (1 - k\dot{m}_{out}\Delta t/m_s)p_s \qquad (24)$$

$$T_{s3} = (1 - (k-1)\dot{m}_{out}\Delta t/m_s)T_s. \qquad (25)$$

Similar to (17) and (18), (24) and (25) are linear (bilinear) equations when used in a one-step (multi-step) optimization problem.

### C. Charging Process Considering Heat Transfer

In Sections II-A and II-B, the heat transfer between the cavern air and the cavern wall is not considered. However, the heat transfer plays an important role in the variation of the air temperature/pressure in the cavern [28]. Therefore, the heat transfer is considered in this subsection and the following two subsections. In this subsection, the temperature as a function of time is first deduced. The pressure as a function of time is then obtained via the ideal gas law. Last, the temperature/pressure as a function of time is linearized to obtain a bilinear model.

According to [28], [29], the air density $(\rho_{av})$ in the cavern and the cavern wall temperature $(T_{RW})$ can be assumed to be constant and the heat transfer between the cavern air and the cavern wall can be modelled as

$$\frac{dT}{dt} = \frac{h_{eff}}{\rho_{av}c_v}(T_{RW} - T) \qquad (26)$$

where

$$h_{eff} = h_a + h_b|\dot{m}_{in} - \dot{m}_{out}|^{0.8} \qquad (27)$$

According to [28], $h_a$ and $h_b$ can be set to 0.2356 and 0.0149, respectively. Equation (26) can be reformed as

$$T(t) = \int \frac{h_{eff}}{\rho_{av}c_v}(T_{RW} - T)dt \qquad (28)$$

Equation (18) can be written as

$$T_{s2}(t) = T_s\left(1 + (k-2)\frac{\dot{m}_{in}t}{m_s}\right) + a_3(m_s)^{k-2}\dot{m}_{in}\,t. \qquad (29)$$

By substituting (29) into (28), i.e., replacing $T$ on the right-hand side of (28) by the right-hand side of (29), one can obtain

$$T_{s2}^{ht}(t) = \int \frac{h_{eff}}{\rho_{av}c_v}\left(T_{RW} - T_s\left(1 + (k-2)\frac{\dot{m}_{in}t}{m_s}\right) - a_3(m_s)^{k-2}\dot{m}_{in}\,t\right)dt \qquad (30)$$

where superscript 'ht' represents 'heat transfer'. By solving the integral equation (30), one can obtain

$$T_{s2}^{ht}(t) = \frac{h_{eff}}{\rho_{av}c_v}\left(T_{RW}\,t - T_s\left(t + (k-2)\frac{\dot{m}_{in}t^2}{2m_s}\right) - a_3(m_s)^{k-2}\dot{m}_{in}\,t^2/2\right). \qquad (31)$$

Adding (29) and (31) together gives

$$T_{s2}^{a,ht}(t) = T_s\left(1 + (k-2)\frac{\dot{m}_{in}t}{m_s}\right) + a_3(m_s)^{k-2}\dot{m}_{in}\,t + \frac{1}{\rho_{av}c_v}\left(h_a + h_b\dot{m}_{in}^{0.8}\right)\left(T_{RW}\,t - T_s\left(t + (k-2)\frac{\dot{m}_{in}t^2}{2m_s}\right) - a_3(m_s)^{k-2}\dot{m}_{in}\frac{t^2}{2}\right) \qquad (32)$$

where superscript 'a,ht' indicates that both the adiabatic process and heat transfer are considered.

According to the first-order Taylor series approximation [36], i.e., $f(x) = f(x_0) + f'(x_0) \times (x - x_0)$, one can linearize $(m_s)^{k-1}$ and $(m_s)^{k-2}$ at $m_{av0}$ as

$$(m_s)^{k-1} = (m_{av0})^{k-1} + (k-1)(m_{av0})^{k-2}(m_s - m_{av0}) \qquad (33)$$

$$(m_s)^{k-2} = (m_{av0})^{k-2} + (k-2)(m_{av0})^{k-3}(m_s - m_{av0}) \qquad (34)$$

where $m_{av0}$ is a fixed value, i.e., $m_{av0} = \rho_{av}V_s$. The $\dot{m}_{in}^{0.8}$ and $\dot{m}_{in}^{1.8}$ in (32) can be linearized using

$$\dot{m}_{in}^{0.8} = \dot{m}_{in0}^{0.8} + l_1(\dot{m}_{in} - \dot{m}_{in0}) \qquad (35)$$

$$\dot{m}_{in}^{1.8} = \dot{m}_{in0}^{1.8} + l_2(\dot{m}_{in} - \dot{m}_{in0}) \qquad (36)$$

where $l_1 = \frac{\dot{m}_{in1}^{0.8} - \dot{m}_{in0}^{0.8}}{\dot{m}_{in1} - \dot{m}_{in0}}$, $l_2 = \frac{\dot{m}_{in1}^{1.8} - \dot{m}_{in0}^{1.8}}{\dot{m}_{in1} - \dot{m}_{in0}}$, and the setting of $\dot{m}_{in1}$ and $\dot{m}_{in0}$ will be given in the beginning of Section IV. By using (33)-(36), equation (32) can then be converted to be a bilinear form, i.e., (37).

$$m_s T_{s2}^{a,ht}(t) = T_s m_s + c_2 T_s \dot{m}_{in} + c_3 m_s \dot{m}_{in} + c_4 \dot{m}_{in} + c_5 T_s + c_6 m_s + c_7 \qquad (37)$$

where $c_2$-$c_7$ are constant coefficients given in the Appendix. Equation (37) involves four variables, i.e., $m_s$, $\dot{m}_{in}$, $T_{s2}^{a,ht}(t)$, and $T_s$ and the first two terms are bilinear terms.

Equation (37) represents the change of the temperature during the charging process as a function of time and the charging mass flow rate $\dot{m}_{in}$, where both the adiabatic process and the heat transfer process are considered.

According to the ideal gas law, one can obtain

$$p_{s2}^{a,ht}(t) = (m_s + \dot{m}_{in}t)RT_{s2}^{a,ht}(t)/V_s \qquad (38)$$

which can be expanded to (39) by substituting (32) therein.

$$p_{s2}^{a,ht}(t) = \frac{(m_s + \dot{m}_{in}t)RT_s}{V_s}\left(1 + (k-2)\frac{\dot{m}_{in}t}{m_s}\right)$$
$$+ \frac{(m_s + \dot{m}_{in}t)R}{V_s}a_3(m_s)^{k-2}\dot{m}_{in}\,t$$
$$+ \frac{(m_s + \dot{m}_{in}t)R}{V_s\rho_{av}c_v}\left(h_a + h_b\dot{m}_{in}^{0.8}\right)\left(T_{RW}\,t - T_s t - T_s(k-2)\frac{\dot{m}_{in}t^2}{2m_s}\right)$$
$$- \frac{(m_s + \dot{m}_{in}t)R}{V_s\rho_{av}c_v}\left(h_a + h_b\dot{m}_{in}^{0.8}\right)\left(a_3(m_s)^{k-2}\dot{m}_{in}\frac{t^2}{2}\right) \qquad (39)$$

The four terms in (39) are each on a separate line. The second term of (39) can be replaced by $m_s Ra_3(m_s)^{k-2}\dot{m}_{in}\,t/V_s$ because $Ra_3/V_s$ is small and $\dot{m}_{in}t$ is much smaller than $m_s$. The last term of (39) can be ignored because the values of both $\left(h_a + h_b\dot{m}_{in}^{0.8}\right)/\rho_{av}c_v$ and $Ra_3/V_s$ are small.

Again, according to the first-order Taylor series approximation [36], i.e., $f(x) = f(x_0) + f'(x_0) \times (x - x_0)$, one can linearize $m_s^k$ at $m_{av0}$:

$$m_s^k = m_{av0}^k + km_{av0}^{k-1}(m_s - m_{av0}) \qquad (40)$$

Now, by using (40), (39) can be reformed into a bilinear form:

$$m_s p_{s2}^{a,ht}(t) = p_s m_s + c_8 p_s \dot{m}_{in} + c_9 m_s \dot{m}_{in} + c_{10}\dot{m}_{in} + c_{11}m_s + c_{12}p_s + c_{13} \qquad (41)$$

where $c_8$-$c_{13}$ are constant coefficients as given in the Appendix. Equation (41) involves four variables, i.e., $m_s$, $\dot{m}_{in}$, $p_{s2}^{a,ht}(t)$, and $p_s$, and the first four terms are bilinear terms.

### D. Discharging Process Considering Heat Transfer

In this subsection, the temperature as a function of time is first deduced. The pressure as a function of time is then obtained via the ideal gas law. Last, the temperature/pressure as a function of time is linearized to obtain a bilinear model. Equation (25) can be written as

$$T(t) = T_s - (k-1)T_s \frac{\dot{m}_{out}}{m_s} t. \tag{42}$$

By substituting (42) into (28), i.e., replacing $T$ on the right-hand side of (28) by the right-hand side of (42), one can obtain

$$T_{s3}^{ht}(t) = \int \frac{h_{eff}}{\rho_{av}c_v} \left( T_{RW} - T_s + (k-1)T_s \frac{\dot{m}_{out}}{m_s} t \right) dt. \tag{43}$$

By solving the integral equation (43), one can obtain

$$T_{s3}^{ht}(t) = \frac{h_{eff}}{\rho_{av}c_v} \left( (T_{RW} - T_s)t + (k-1)T_s \frac{\dot{m}_{out}}{2m_s} t^2 \right). \tag{44}$$

Adding (42) and (44) together gives

$$T_{s3}^{a,ht}(t) = T_s - (k-1)T_s \frac{\dot{m}_{out}}{m_s} t + \frac{h_{eff}}{\rho_{av}c_v} \left( (T_{RW} - T_s)t + (k-1)T_s \frac{\dot{m}_{out}}{2m_s} t^2 \right). \tag{45}$$

Equation (45) represents the change in temperature during the discharging process as a function of time and discharging mass flow rate $\dot{m}_{out}$, where both the adiabatic process and the heat transfer process are considered. Equation (45) can be rewritten as

$$m_s T_{s3}^{a,ht}(t) = m_s T_s - (k-1)T_s \dot{m}_{out} t + \frac{V_s}{c_v}(T_{RW} - T_s)t \left( h_a + h_b \dot{m}_{out}^{0.8} \right) + \frac{1}{2\rho_{av}c_v}(k-1)T_s t^2 (h_a \dot{m}_{out} + h_b \dot{m}_{out}^{1.8}). \tag{46}$$

Similarly, $\dot{m}_{out}^{0.8}$ and $\dot{m}_{out}^{1.8}$ can be linearized using

$$\dot{m}_{out}^{0.8} = \dot{m}_{out0}^{0.8} + l_3(\dot{m}_{out} - \dot{m}_{out0}) \tag{47}$$

$$\dot{m}_{out}^{1.8} = \dot{m}_{out0}^{1.8} + l_4(\dot{m}_{out} - \dot{m}_{out0}) \tag{48}$$

where $l_3 = \frac{\dot{m}_{out1}^{0.8} - \dot{m}_{out0}^{0.8}}{\dot{m}_{out1} - \dot{m}_{out0}}$, $l_4 = \frac{\dot{m}_{out1}^{1.8} - \dot{m}_{out0}^{1.8}}{\dot{m}_{out1} - \dot{m}_{out0}}$, and the setting of $\dot{m}_{out1}$ and $\dot{m}_{out0}$ will be given at the beginning of Section IV. Now, (46) can be reformulated into a bilinear form:

$$m_s T_{s3}^{a,ht}(t) = m_s T_s + c_{14}T_s \dot{m}_{out} + c_{15}\dot{m}_{out} + c_{16}T_s + c_{17} \tag{49}$$

where $c_{14}$-$c_{17}$ are constant coefficients as given in the Appendix. Equation (49) involves four variables, i.e., $m_s$, $\dot{m}_{out}$, $T_{s3}^{a,ht}(t)$, and $T_s$, and the first three terms are bilinear terms. According to the ideal gas law, one can obtain

$$p_{s3}^{a,ht}(t) = (m_s - \dot{m}_{out}t)RT_{s3}^{a,ht}(t)/V_s \tag{50}$$

which can be expanded as follows by substituting (45) therein:

$$p_{s3}^{a,ht}(t) = \frac{(m_s - \dot{m}_{out}t)RT_s}{V_s} - \frac{(m_s - \dot{m}_{out}t)R}{V_s}(k-1)T_s \frac{\dot{m}_{out}}{m_s} t + \frac{h_{eff}}{\rho_{av}c_v} \left( \frac{(m_s - \dot{m}_{out}t)R}{V_s}(T_{RW} - T_s)t + \frac{(m_s - \dot{m}_{out}t)R}{V_s}(k-1)T_s \frac{\dot{m}_{out}}{2m_s}t^2 \right) \tag{51}$$

Note that $(m_s - \dot{m}_{out}t)\frac{\dot{m}_{out}}{m_s}$ in the third term of (51) can be replaced by $m_s \frac{\dot{m}_{out}}{m_s}$ because $\dot{m}_{out}t$ is much smaller than $m_s$. Equation (51) can then be reformed as

$$m_s p_{s3}^{a,ht}(t) = (m_s - k\dot{m}_{out}t)p_s + \frac{h_{eff}R}{c_v}((m_s - \frac{\dot{m}_{out}t}{2})(T_{RW} - T_s)t + 0.5(k-1)T_s \dot{m}_{out}t^2) \tag{52}$$

Comparing (52) with (24), we know that the term $(m_s - k\dot{m}_{out}t)p_s$ in (52) represents the adiabatic process inside the cavern and the terms in the second line of (52) are associated with the heat transfer between the cavern air and the cavern wall. Equation (52) can be further reformulated into the following bilinear form:

$$m_s p_{s3}^{a,ht}(t) = p_s m_s + c_{18}p_s \dot{m}_{out} + c_{19}T_s m_s + c_{20}T_s \dot{m}_{out} + c_{21}\dot{m}_{out}\dot{m}_{out} + c_{22}m_s \dot{m}_{out} + c_{23}\dot{m}_{out} + c_{24}m_s + c_{25}p_s + c_{26}T_s \tag{53}$$

where $c_{18}$-$c_{26}$ are constant coefficients as given in the Appendix. Equation (53) involves five variables, i.e., $m_s$, $\dot{m}_{out}$, $p_{s3}^{a,ht}(t)$, $p_s$, and $T_s$, and the first seven terms are bilinear terms.

### E. Idle Process Considering Heat Transfer

When idling, i.e., when neither charging nor discharging, heat transfer occurs between the cavern air and the cavern wall if a temperature difference exists. By solving the integral equation (28), the change of air temperature in the cavern in the idle process can be expressed as

$$T_{s4}^{ht}(t) = (T_s - T_{RW})e^{-\frac{h_{eff}}{\rho_{av}c_v}t} + T_{RW} \tag{54}$$

where $T_s$ is the initial air temperature of the cavern in the idle process. According to the ideal gas law, one can obtain

$$p_{s4}^{ht}(t) = m_s RT_{s4}^{ht}(t)/V_s \tag{55}$$

which can be expanded into (56) by substituting (54) therein:

$$p_{s4}^{ht}(t) = m_s R(T_s - T_{RW})e^{-\frac{h_{eff}}{\rho_{av}c_v}t}/V_s + m_s RT_{RW}/V_s \tag{56}$$

Equation (56) can be reformed as

$$p_{s4}^{ht}(t) = p_s e^{-\frac{h_{eff}}{\rho_{av}c_v}t} + m_s RT_{RW}(1 - e^{-\frac{h_{eff}}{\rho_{av}c_v}t})/V_s \tag{57}$$

The $e^{-\frac{h_{eff}}{\rho_{av}c_v}t}$ in (57) can be expressed as $e^{-\frac{h_{eff}V_s}{m_{av}c_v}t}$, which can be linearized as follows:

$$e^{-\frac{h_c A_c}{m_{av}c_v}t} = e^{-a_4} + \frac{a_4}{m_{av0}}e^{-a_4}(m_s - m_{av0}) \tag{58}$$

where $a_4 = \frac{h_{eff}V_s t}{m_{av0}c_v}$. By substituting (58) into (54) and (57), one can obtain

$$T_{s4}^{\text{ht}}(t) = (T_s - T_{RW})(e^{-a_4} + a_4 e^{-a_4}(m_s - m_{av0})/m_{av0}) + T_{RW} \quad (59)$$

$$p_{s4}^{\text{ht}}(t) = p_s(e^{-a_4} + a_4 e^{-a_4}(m_s - m_{av0})/m_{av0})$$
$$+ \frac{m_s R T_{RW}}{V_s}\left(1 - e^{-a_4} - \frac{a_4}{m_{av0}}e^{-a_4}(m_s - m_{av0})\right) \quad (60)$$

Equations (59) and (60) can be reformulated into the following bilinear forms:

$$T_{s4}^{\text{ht}}(t) = c_{27}m_s T_s + c_{28}T_s + c_{29}m_s + c_{30} \quad (61)$$

$$p_{s4}^{\text{ht}}(t) = c_{27}p_s m_s + c_{31}m_s m_s + c_{28}p_s + c_{32}m_s \quad (62)$$

where $c_{27} - c_{32}$ are constant coefficients as given in the Appendix. Equation (61) involves three variables, i.e., $m_s$, $T_{s4}^{\text{ht}}(t)$, and $T_s$, and the $2^{\text{nd}}$ term is the only bilinear term. Equation (62) involves three variables, i.e., $m_s$, $p_{s4}^{\text{ht}}(t)$, and $p_s$, and the $2^{\text{nd}}$ and $3^{\text{rd}}$ terms are bilinear.

This completes the deduction of the bilinear cavern model. In summary, the bilinear cavern model includes (37) and (41) for the charging process, (49) and (53) for the discharging process, and (61) and (62) for the idle process.

## III. Self-scheduling of Compressed Air Energy Storage in the Day-Ahead Electricity Market

### A. Self-scheduling Problem Using the Bilinear Cavern Model

In the SSP of CAES, the day-ahead electricity market is considered and the CAES plant is assumed to be a price taker (i.e., the CAES plant does not affect the electricity market price). The objective of SSP is to maximize the arbitrage revenue obtained from selling electricity to and buying electricity from the day-ahead electricity market, i.e., selling electricity (in the discharging process) to the market in high-electricity-price periods and buying electricity (in the charging process) from the market in lower-electricity-price periods. This implies that the aim of SSP is to decide when to charge and discharge subjected to physical constraints of the CAES plant, including the minimum/maximum pressure range of the storage cavern air.

The objective function of the SSP can be expressed by (63), which represents the net profit obtained by the CAES plant from the electricity market. Note that the operational costs of charging and discharging power in the CAES are considered in (63). Equation (64) represents the relationship between the mass flow rate in ($\dot{m}_{\text{in}}^t$) and charging power ($P_{\text{ch}}^t$) and (65) represents the relationship between the mass flow rate out ($\dot{m}_{\text{out}}^t$) and discharging power ($P_{\text{dch}}^t$) [12]. Equation (66) represents the lower and upper bounds of the air pressure in the cavern. Equation (67) ensures that the charging and discharging processes do not occur at the same time. Equations (68) and (69) describe the ranges of charging and discharging power and their relationships with the charging and discharging status indicators, i.e., $\alpha^t$ and $\beta^t$, respectively.

maximize: $f_{\text{obj}} = \sum_{t \in \Omega_{T1}} ((\tau_t - C_{\text{dch}}^t)P_{\text{dch}}^t - (\tau_t + C_{\text{ch}}^t)P_{\text{ch}}^t)\frac{\Delta t}{3600}$ (63)

$$\dot{m}_{\text{in}}^t = c_{\text{Ain}}P_{\text{ch}}^t, \quad \forall t \in \Omega_{T1} \quad (64)$$

$$\dot{m}_{\text{out}}^t = c_{\text{Aout}}P_{\text{dch}}^t, \quad \forall t \in \Omega_{T1} \quad (65)$$

$$p_{\text{min}} \le p_s^t \le p_{\text{max}}, \quad \forall t \in \Omega_{T1} \quad (66)$$

$$\alpha^t + \beta^t \le 1, \quad \forall t \in \Omega_{T1} \quad (67)$$

$$\alpha^t P_{\text{ch}}^{\text{min}} \le P_{\text{ch}}^t \le \alpha^t P_{\text{ch}}^{\text{max}}, \quad \forall t \in \Omega_{T1} \quad (68)$$

$$\beta^t P_{\text{dch}}^{\text{min}} \le P_{\text{dch}}^t \le \beta^t P_{\text{dch}}^{\text{max}}, \quad \forall t \in \Omega_{T1} \quad (69)$$

where the values of the coefficients $c_{\text{Ain}}$ and $c_{\text{Aout}}$ are adopted from [26].

The bilinear cavern model presented in the previous section, i.e., (37), (41), (49), (53), (61) and (62), is re-written below using superscript $t$ and $(t+1)$ to represent the indices of time steps.

$$m_s^t T_{s,\text{ch}}^{(t+1)} = T_s^t m_s^t + c_2 T_s^t \dot{m}_{\text{in}}^{(t+1)} + c_3 m_s^t \dot{m}_{\text{in}}^{(t+1)} + c_4 \dot{m}_{\text{in}}^{(t+1)} + c_5 T_s^t + c_6 m_s^t + c_7, \quad \forall t \in \Omega_{T0} \quad (70)$$

$$m_s^t p_{s,\text{ch}}^{(t+1)} = p_s^t m_s^t + c_8 p_s^t \dot{m}_{\text{in}}^{(t+1)} + c_9 m_s^t \dot{m}_{\text{in}}^{(t+1)} + c_{10}\dot{m}_{\text{in}}^{(t+1)} + c_{11}m_s^t + c_{12}p_s^t + c_{13}, \quad \forall t \in \Omega_{T0} \quad (71)$$

$$m_s^t T_{s,\text{dch}}^{(t+1)} = m_s^t T_s^t + c_{14}T_s^t \dot{m}_{\text{out}}^{(t+1)} + c_{15}\dot{m}_{\text{out}}^{(t+1)} + c_{16}T_s^t + c_{17}, \quad \forall t \in \Omega_{T0} \quad (72)$$

$$m_s^t p_{s,\text{dch}}^{(t+1)} = p_s^t m_s^t + c_{18}p_s^t \dot{m}_{\text{out}}^{(t+1)} + c_{19}T_s^t m_s^t + c_{20}T_s^t \dot{m}_{\text{out}}^{(t+1)} + c_{21}\dot{m}_{\text{out}}^{(t+1)}\dot{m}_{\text{out}}^{(t+1)} + c_{22}m_s^t \dot{m}_{\text{out}}^{(t+1)} + c_{23}\dot{m}_{\text{out}}^{(t+1)} + c_{24}m_s^t + c_{25}p_s^t + c_{26}T_s^t, \quad \forall t \in \Omega_{T0} \quad (73)$$

$$T_{s,\text{idl}}^{(t+1)} = c_{27}m_s^t T_s^t + c_{28}T_s^t + c_{29}m_s^t + c_{30}, \quad \forall t \in \Omega_{T0} \quad (74)$$

$$p_{s,\text{idl}}^{(t+1)} = c_{27}p_s^t m_s^t + c_{31}m_s^t m_s^t + c_{28}p_s^t + c_{32}m_s^t, \quad \forall t \in \Omega_{T0} \quad (75)$$

The $T_{s,\text{ch}}^{(t+1)}$ ($T_{s,\text{dch}}^{(t+1)}$, $T_{s,\text{idl}}^{(t+1)}$) represents the temperature of the cavern air if the $t$th step is a charging (discharging, idle) process. The temperature at the beginning of the $(t+1)$th step, i.e., $T_s^{(t+1)}$, is equal to one of $T_{s,\text{ch}}^{(t+1)}$, $T_{s,\text{dch}}^{(t+1)}$, and $T_{s,\text{idl}}^{(t+1)}$ according to the values of $\alpha^{(t+1)}$ and $\beta^{(t+1)}$, which is modelled in (76). The relationship between the pressure at the beginning of the $(t+1)$th step, i.e., $p_s^{(t+1)}$, and $p_{s,\text{ch}}^{(t+1)}$, $p_{s,\text{dch}}^{(t+1)}$, and $p_{s,\text{idl}}^{(t+1)}$ can be similarly modeled as (77). The relationship among the mass of air at two consecutive steps ($m_s^t$ and $m_s^{(t+1)}$) and the charging/discharging status indicators ($\alpha^{(t+1)}$ and $\beta^{(t+1)}$) can be expressed as (78).

$$T_s^{(t+1)} = \alpha^{(t+1)}T_{s,\text{ch}}^{(t+1)} + \beta^{(t+1)}T_{s,\text{dch}}^{(t+1)} + (1 - \alpha^{(t+1)} - \beta^{(t+1)})T_{s,\text{idl}}^{(t+1)}, \quad \forall t \in \Omega_{T0} \quad (76)$$

$$p_s^{(t+1)} = \alpha^{(t+1)}p_{s,\text{ch}}^{(t+1)} + \beta^{(t+1)}p_{s,\text{dch}}^{(t+1)} + (1 - \alpha^{(t+1)} - \beta^{(t+1)})p_{s,\text{idl}}^{(t+1)}, \quad \forall t \in \Omega_{T0} \quad (77)$$

$$m_s^{(t+1)} = m_s^t + \alpha^{(t+1)}\dot{m}_{\text{in}}^{(t+1)}\Delta t - \beta^{(t+1)}\dot{m}_{\text{out}}^{(t+1)}\Delta t, \quad \forall t \in \Omega_{T0} \quad (78)$$

The SSP of CAES using the bilinear model, i.e., (63)-(78), is

 

referred to as Model 1, which is an MIBLP model. The bilinear terms in Model 1 can be expressed as $xy$, which is equal to $(x+y)^2/4 - (x-y)^2/4$. Then, $(x+y)^2$ and $(x-y)^2$ can be piecewise linearized. The detailed procedure of this linearization is available in Section IV-C of [40]. The resulting MILP model is referred to as Model 2.

### B. Flowcharts for Using the Bilinear Cavern Model

The flowcharts for using the bilinear model, i.e., (70)-(78), to calculate the temperature, pressure, and mass of air in the charging, discharging, and idle processes are given in Figs. 4a, 4b, and 4c, respectively. The calculation procedures given in Figs. 4a, 4b, and 4c are recursive but consume very little time because equations (70)-(75) each only have one variable per step that can simply be calculated from the other known values. For example, $T_{s,ch}^{t+1}$ is the only variable in (70) and all the other notations in (70) are known. Flowcharts given in Figs. 4a, 4b, and 4c are used in Sections IV-A to IV-C.

The flowchart for using the bilinear cavern model for an optimization problem is given in Fig. 4d, i.e., using (70)-(78) in an optimization problem. The bilinear cavern model part, i.e., (70)-(78), in an optimization model has $11\,n_t$ continuous variables (i.e., $T_s^t$, $p_s^t$, and $m_s^t$ for $t=1,2,\cdots,n_t$ and $T_{s,ch}^{t+1}$, $p_{s,ch}^{t+1}$, $T_{s,dch}^{t+1}$, $p_{s,dch}^{t+1}$, $T_{s,idl}^{t+1}$, $p_{s,idl}^{t+1}$, $\dot{m}_{in}^{t+1}$, and $\dot{m}_{out}^{t+1}$ for $t=0,1,2,\cdots,(n_t-1)$) and $2n_t$ binary variables (i.e., $\alpha^{t+1}$ and $\beta^{t+1}$ for $t=0,1,2,\cdots,(n_t-1)$). The flowchart given in Fig. 4d is used in Sections IV-D and IV-E. Equations (70)-(78) together with the objective and other constraints, e.g., (63)-(69), form an optimization problem that can be solved by an MINLP solver, further details of which are given in Section IV-D.

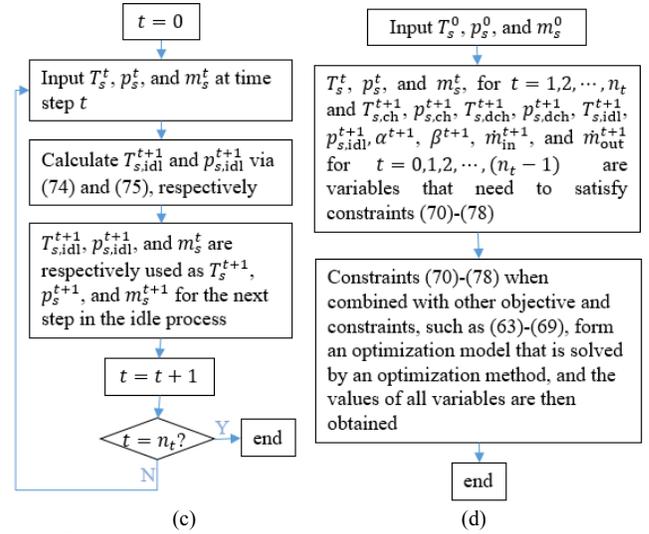

Fig. 4. Flowchart for calculating the temperature, pressure, and mass of air using the bilinear model: (a) in the charging process, (b) in the discharging process, (c) in the idle process; (d) flowchart for using the proposed bilinear model in an optimization model.

### C. Self-scheduling Problem Using Nonlinear and Constant-temperature Cavern Models

For the sake of completeness, the nonlinear cavern model given in [29] is rewritten here, i.e., (79)-(82). The SSP of CAES using the nonlinear cavern model given in [29] consists of (63)-(69), (76), and (78)-(82), and is referred to as Model 3.

$$T_{s,ch}^{(t+1)} = \left(T_s^t + \frac{\dot{m}_{in}^{(t+1)}c_p T_s^t + h_c A_c T_{RW}}{\dot{m}_{in}^{(t+1)}(R-c_p) - h_c A_c}\right) e^{\frac{\dot{m}_{in}^{(t+1)}(R-c_p) - h_c A_c}{V_s \rho_{av} c_v}\Delta t} - \frac{\dot{m}_{in}^{(t+1)}c_p T_s^t + h_c A_c T_{RW}}{\dot{m}_{in}^{(t+1)}(R-c_p) - h_c A_c} \quad (79)$$

$$T_{s,dch}^{(t+1)} = \left(T_s^t + \frac{h_c A_c T_{RW}}{\dot{m}_{out}^{(t+1)}R - h_c A_c}\right) e^{\frac{\dot{m}_{out}^{(t+1)}R - h_c A_c}{V_s \rho_{av} c_v}\Delta t} - \frac{h_c A_c T_{RW}}{\dot{m}_{out}^{(t+1)}R - h_c A_c} \quad (80)$$

$$T_{s,idl}^{(t+1)} = (T_s^t - T_{RW})e^{\frac{-h_c A_c}{V_s \rho_{av} c_v}\Delta t} + T_{RW} \quad (81)$$

$$p_s^{(t+1)} = T_s^{(t+1)} m_s^{(t+1)} R / V_s \quad (82)$$

For the sake of comparison, the constant-temperature cavern model is also provided here, i.e., (83). The SSP of CAES using the constant-temperature model consists of (63)-(69), (76), and (83), which is referred to as Model 4. Models 1-4 will be compared in Section IV-E.

$$T_s^{(t+1)} = T_s^t \quad (83)$$

## IV. SIMULATION

In this paper, the parameters for the Huntorf CAES plant are used for the calculations. The Huntorf CAES plant features two caverns with volumes of 141,000 and 169,000 m³, respectively. Note that the maximum mass flow rate in the charging process ($\dot{m}_{in}$) is 108 kg/s for the whole plant and 49.12 kg/s for the first cavern, which is calculated from $108 \times \frac{141000}{141000+169000}$. Similarly, the maximum mass flow rate in the discharging process ($\dot{m}_{out}$) is 417 kg/s for the whole plant and 189.67 kg/s for the first cavern, which is calculated from $417 \times \frac{141000}{141000+169000}$. In this paper, the first cavern is used for calculations. The $\dot{m}_{in1}$ and $\dot{m}_{in0}$ used in (35) and (36) are set to 49.12 and 22.74 kg/s, respectively. The $\dot{m}_{out1}$ and $\dot{m}_{out0}$ used in (47) and (48) are set to 189.67 and 90.97 kg/s, respectively. The temperature at the outlet of high-pressure compressor is high (about 230 °C), but drops to 50 °C after going through an aftercooler that is located between the high-pressure compressor and the cavern, i.e., the temperature of air injected into the caverns is equal to 50 °C. The other parameters for the Huntorf CAES plant are given in Table I [28], [29].

TABLE I
PARAMETERS FOR THE HUNTORF CAES PLANT.

| $A_c$ | $c_v$ | $k$ | $p_{in}$ | $C_{ch}^t$ |
|---|---|---|---|---|
| 25,000 m² | 718.3 J/(kg K) | 1.4 | 66 bar | 3 $/MWh |
| $R$ | $V_s$ | $T_{RW}$ | $T_{in}$ | $C_{dch}^t$ |
| 286.7 J/(kg K) | 141,000 m³ | 40 °C | 50 °C | 3 $/MWh |

Three specific processes are defined as follows and used to verify the accuracy of the proposed bilinear model:

- Charging process: Set the initial pressure (temperature) of the air in the cavern to 46 bar (20 °C). Charge the first cavern (141,000 m³) continuously for 16 hours at the maximum mass flow rate, i.e., $\dot{m}_{in}$ =49.12 kg/s.
- Discharging process: Set the initial pressure (temperature) of the air in the cavern to 66 bar (40 °C). Discharge the cavern continuously for 4 hours at the maximum mass flow rate, i.e., $\dot{m}_{out}$ =189.67 kg/s.
- Idle process: Set the initial pressure (temperature) of the air in the cavern to 60 bar (45 °C). Let the cavern be in the idle process for 16 hours.

Reference [29] compares several existing CAES models with the measured data from the Huntorf CAES plant. The analytical model in [29] is accurate and simpler than other existing analytical models. Thus, the analytical model in [29] is used as a benchmark model in this section to verify the accuracy of the proposed bilinear cavern model.

### A. Model Verification - Comparison With Accurate Model

In this subsection, the time interval is set to 1 second, i.e., $\Delta t$ is equal to 1 second in (70)-(75). The pressure and temperature for each time interval of the charging (discharging, idle) process obtained from both the proposed bilinear model and the analytical model in [29] are plotted in Fig. 5 (Fig. 6, Fig. 7). Note that the difference between the results of the two models would be difficult to observe if the results for each second were shown in Figs. 5-7; therefore, the results for every 1000 (250, 1000) seconds are shown in Fig. 5 (Fig. 6, Fig. 7). Figs. 5-7 show that the pressure/temperature results obtained from the proposed bilinear model and the analytical model are quite close to one another.

The mean absolute percentage errors (MAPEs) between the results, in terms of both pressure and temperature, obtained from the bilinear model and the analytical model during the charging, discharging, and idle processes are tabulated in Table II. The last column of Table II shows that the idle part of the bilinear model, i.e., (74) and (75), is almost as accurate as the analytical model. The 2$^{nd}$ and 3$^{rd}$ columns of Table II show that the inaccuracy of the charging/discharging parts of the bilinear model, i.e., (70)-(73), is around 0.11%, which is quite small. This verifies the accuracy of the bilinear model for the given initial temperature/pressure and mass flow rate.

TABLE II
THE MAPE BETWEEN THE RESULTS OBTAINED BY THE BILINEAR MODEL AND THE ANALYTICAL MODEL GIVEN IN [29] IN EACH OF THE THREE PROCESSES.

|  | Charging process | Discharging process | Idle process |
|---|---|---|---|
| Pressure | 0.0011 | 0.0011 | $1.12 \times 10^{-5}$ |
| Temperature | 0.0011 | 0.0011 | $1.12 \times 10^{-5}$ |

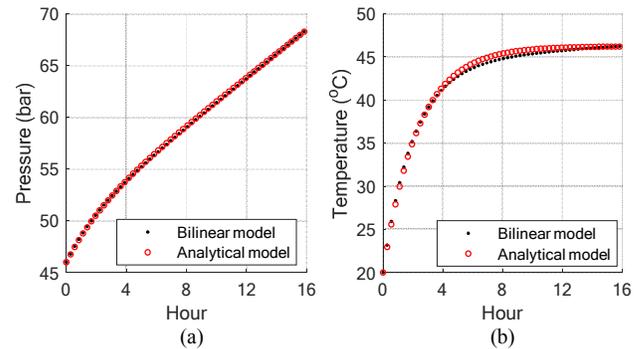

Fig. 5. Results obtained by the proposed bilinear model and the analytical model in [29] during the charging process: a) pressure, b) temperature.

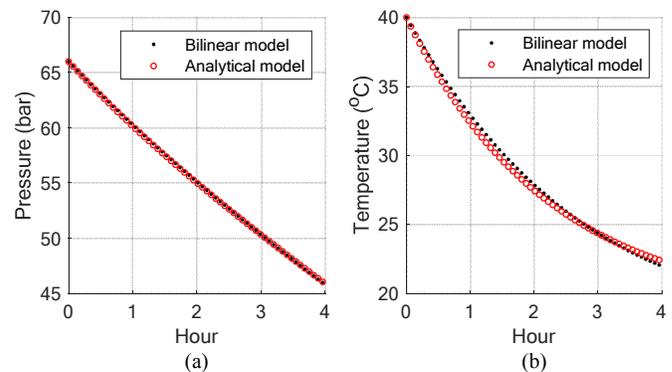

Fig. 6. Results obtained by the proposed bilinear model and the analytical model in [29] during the discharging process: a) pressure, b) temperature.

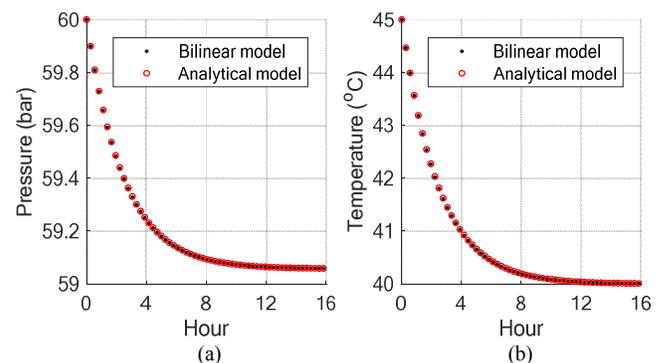

Fig. 7. Results obtained by the proposed bilinear model and the analytical model in [29] during the idle process: a) pressure, b) temperature.

To comprehensively evaluate the accuracy of the bilinear model, different settings of initial temperatures/pressures and mass flow rates are used in the charging/discharging/idle process. Different settings for the charging (discharging, idle) process are shown in rows C1-C6 (D1-D7, I1-I4) of Table III. When the pressure is larger (smaller) than 46 bar, the CAES is in normal (emergency) operating mode. Fig. 5 (Fig. 6) shows that the cavern temperature is relatively high (low) when the pressure is high (low). Therefore, the initial temperatures are set to 20 and 35 °C (50 and 35 °C) in the different settings for the charging (discharging) process as its initial pressure is relatively low (high). Both large and small mass flow rates are investigated in the normal operating mode as given in C1-C4 and D1-D4 where the large and small values are set to the maximum mass flow rate and one tenth of it, respectively. To avoid a lengthy table, 1) only results that tend to have a large error in the emergency model are listed in Table III, i.e., low mass flow rate (associated with low charging power) and low initial temperature as indicated by setting C1-C4, and low mass flow rate (associated with low discharging power) and high initial temperature as indicated by settings D1-D4; and 2) not all the results of different settings for the idle process are listed because all of the different settings have very small errors, i.e., the MAPE is less than 5E-5. Note that 'E-n' and 'E+n' are used in this paper to represent '$\times 10^{-n}$' and '$\times 10^{n}$', respectively.

For each setting, the MAPE and mean absolute error (MAE) between the results (including both pressure and temperature) obtained by the bilinear and the analytical models is given in the last four columns of Table III. The largest error in different settings for the charging process occurs at C2 with low initial temperature and low mass flow rate; the MAPE and MAE for pressure (temperature) are 0.77% and 0.38 bar (0.77% and 2.38 °C), respectively. The largest error in different settings for the discharging process occurs at D5 with high initial temperature and low mass flow rate; the MAPE and MAE for pressure (temperature) are 0.82% and 0.36 bar (0.82% and 2.58 °C), respectively. Note that the MAPE and MAE are usually less than the worst-case values given above. The errors in different settings for the idle process are very small; the MAPE for both pressure and temperature is less than 5E-5. This indicates that the error between the proposed bilinear cavern model and the accurate analytical model is small under a variety of circumstances.

TABLE III
DIFFERENT SETTINGS OF INITIAL PRESSURES AND TEMPERATURES AND CHARGING/DISCHARGING POWER AND THE CORRESPONDING MAPE AND MAE BETWEEN RESULTS OBTAINED BY THE BILINEAR CAVERN MODEL AND THE ANALYTICAL MODEL GIVEN IN [29] IN EACH OF THE THREE PROCESSES

| Setting | Initial pressure (bar) | Initial temperature (°C) | Mass flow rate (kg/s) | Pressure | | Temperature | |
|---|---|---|---|---|---|---|---|
| | | | | MAPE | MAE (bar) | MAPE | MAE (°C) |
| C1 | 46 | 20 | 49.12 | 0.0011 | 0.07 | 0.0011 | 0.35 |
| C2 | 46 | 20 | 4.912 | 0.0077 | 0.38 | 0.0077 | 2.38 |
| C3 | 46 | 35 | 49.12 | 0.0021 | 0.11 | 0.0020 | 0.64 |
| C4 | 46 | 35 | 4.912 | 0.0017 | 0.08 | 0.0017 | 0.53 |
| C5 | 30 | 20 | 4.912 | 0.0060 | 0.19 | 0.0060 | 1.84 |
| C6 | 5 | 20 | 4.912 | 0.0047 | 0.03 | 0.0047 | 1.46 |
| D1 | 66 | 50 | 189.67 | 0.0018 | 0.10 | 0.0018 | 0.57 |
| D2 | 66 | 50 | 18.967 | 0.0078 | 0.50 | 0.0078 | 2.47 |
| D3 | 66 | 35 | 189.67 | 0.0008 | 0.04 | 0.0008 | 0.24 |
| D4 | 66 | 35 | 18.967 | 0.0056 | 0.37 | 0.0056 | 1.75 |
| D5 | 46 | 50 | 18.967 | 0.0082 | 0.36 | 0.0082 | 2.58 |
| D6 | 30 | 50 | 18.967 | 0.0071 | 0.20 | 0.0071 | 2.22 |
| D7 | 5 | 50 | 18.967 | 0.012 | 0.047 | 0.012 | 3.82 |
| I1 | 46 | 20 | ---- | 3.9E-5 | 1.8E-3 | 3.9E-5 | 1.2E-2 |
| I2 | 5 | 20 | ---- | 4.2E-6 | 2.2E-5 | 4.2E-6 | 1.3E-3 |
| I3 | 66 | 50 | ---- | 2.4E-5 | 1.6E-3 | 2.4E-5 | 7.7E-3 |
| I4 | 5 | 50 | ---- | 1.8E-6 | 9.1E-6 | 1.8E-6 | 5.9E-4 |

### B. Model Verification - Comparison With Field-Measured Data

To further verify the accuracy of the proposed bilinear model, field-measured pressure and temperature data during a discharging process from [28] and a 10-stage charging/discharging/idle process from [29] are used and referred to as field-measured data I and II, respectively. The pressure and temperature results from both the proposed bilinear model and field-measured data I are plotted in Figs. 8b and 8c, respectively. These figures show the pressures/temperatures obtained from the proposed bilinear model are close to field-measured data I. Figs. 8b and 8c are both divided by a dashed line into two parts: the left-hand (right-hand) part corresponds to pressure being higher (lower) than 46 bar. For the left-hand (right-hand) part, the MAPE and MAE between the results obtained by the bilinear model and field-measured data I are 0.47% and 0.27 bar (5.85% and 1.94 bar) for the pressure and 0.19% and 0.56 °C (0.23% and 0.64 °C) for the temperature, where the unit of temperature is set to

Kelvin when calculating the MAPE (Kelvin is the unit of temperature for all calculations involved in the bilinear cavern model). Note that the cavern pressure being higher (lower) than 46 bar indicates normal (emergency) operating mode. The CAES usually runs in normal operating mode. Therefore, the proposed bilinear model is quite accurate when compared to field-measured data I in the normal operating mode of CAES.

For field-measured data II, the air mass flow rates in the 10 stages are given in Fig. 9a where negative (positive) values represent the discharging (charging) process and the zero value represents the idle process. In Fig. 9a, the 10 stages are divided by nine vertical dashed lines. In stage 1, the air mass flow rate linearly decreases from -125 to -175 kg/s and then increases to -125 kg/s. In stage 5, the CAES is discharging at a fixed air mass flow rate of -175 kg/s. In stages 3, 7, and 9, the CAES is charging at a fixed air mass flow rate of 50 kg/s. In stages 2, 4, 6, 8, and 10, the CAES is in the idle stage, i.e., the air mass flow rate is zero. The pressures (temperatures) at each stage obtained from the bilinear model, analytical model, and field-measured data II are plotted in Fig. 9b (Fig. 9c). Fig. 9b shows the

                                                                     

pressures obtained from both the bilinear model and the analytical model are quite close to field-measured data II except for stage 2. The MAPE and MAE between the pressure obtained by the bilinear (analytical) model and field-measured data II are 1.01% and 0.54 bar (0.97% and 0.51 bar), respectively. In Fig. 9c, the temperature obtained from the bilinear model is closer to field-measured data II than the analytical model. The MAPE and MAE between the temperature obtained by the bilinear (analytical) model and field-measured data II are 3.16% and 1.18 °C (3.88% and 1.47 °C), respectively.

In summary, the pressure (temperature) result obtained from the bilinear cavern model can well match the pressure (temperature) from both field-measured data I and II with a relative error of 0.47 and 1.01% (0.19 and 3.16%), respectively, in the normal operating mode of CAES. This comparison with both an accurate model and two sets of field-measured data verifies the accuracy of the proposed bilinear cavern model.

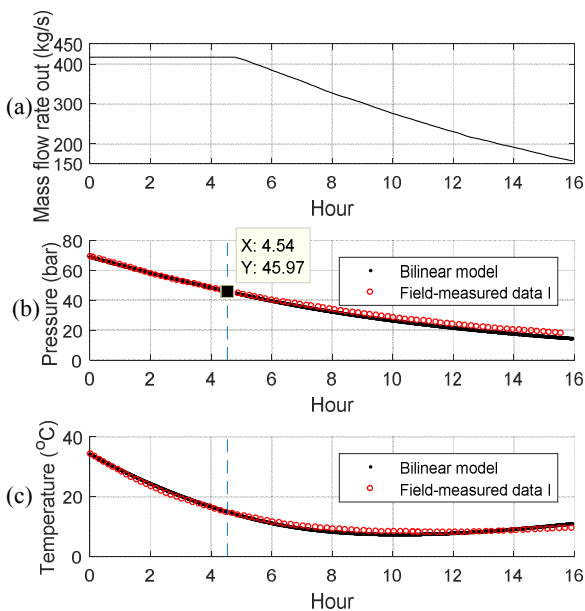

Fig. 8. Mass flow rate out of cavern (a) and comparison between the results obtained by the bilinear cavern model and field-measured data I from [28] for pressure (b) and temperature (c).

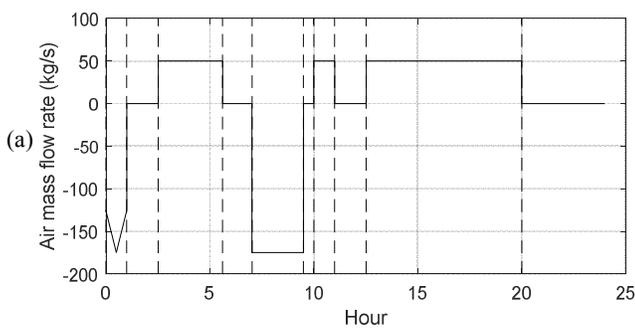

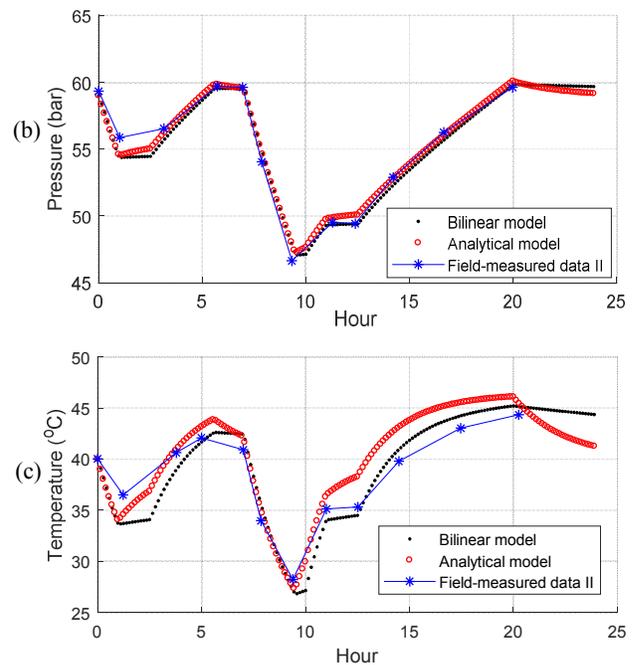

Fig. 9. Mass flow rate into cavern (a) and comparison between the results obtained by the bilinear cavern model, analytical cavern model, and field-measured data II obtained from [29] for pressure (b) and temperature (c).

### C.  Impact of Heat Transfer and Temperature

To observe the impact of heat transfer, the results obtained from the bilinear model with and without considering heat transfer in the charging (discharging) process are plotted in Fig. 10 (Fig. 11). In Fig. 10b, the slope of the left-hand part and the right-hand part of the dotted line (results of the model with heat transfer) is higher and lower, respectively, than the circled line (results of the model without heat transfer). The reason is that when the cavern air temperature is lower (higher) than the cavern wall temperature, i.e., 40 °C, the heat transfer provides heat to (absorbs heat from) the cavern air and therefore the temperature obtained by the model with heat transfer increases faster (slower) than the model without heat transfer. According to the ideal gas law, i.e., $pV = mRT$, for the same $V$, $m$, and $R$, higher (lower) temperature corresponds to higher (lower) pressure and therefore the relationship between the dotted and circled lines in Fig. 10a is similar to Fig. 10b. In Fig. 11, the dotted line has a smaller slope than the circled line. The reason is that the heat transfer provides heat to the cavern air and therefore the temperature decrease associated with the model with heat transfer is smaller than the model without heat transfer. Therefore, heat transfer clearly has a significant impact on the temperature and pressure variation and, consequently, it is important to consider heat transfer in the cavern model.

The results obtained from the constant-temperature model in the charging (discharging) process are also plotted in Fig. 10 (Fig. 11). Obviously, the pressure and temperature obtained from the constant-temperature model are quite different from those obtained with the bilinear model. Considering that the accuracy of the bilinear model was verified in Section V-A, Figs. 10 and 11 indicate that the constant-temperature cavern model is inaccurate. Therefore, it is necessary to use an accurate cavern model instead of the constant-temperature cavern model, especially in a real application problem.

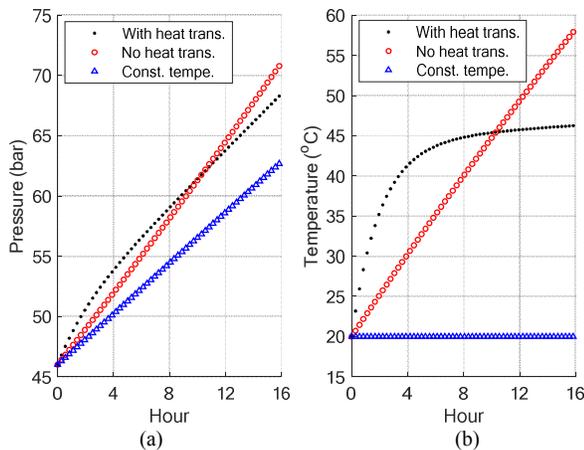

Fig. 10. Results obtained from the bilinear cavern model (with and without considering heat transfer) and the constant-temperature cavern model in the charging process: a) pressure of the air in the cavern and b) temperature of the air in the cavern.

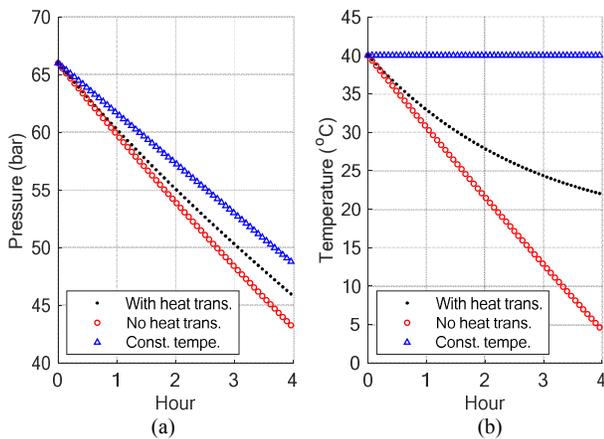

Fig. 11. Results obtained from the bilinear cavern model (with and without considering heat transfer) and the constant-temperature cavern model in the discharging process: (a) pressure of the air in the cavern and (b) temperature of the air in the cavern.

## D. Different Time Intervals

In power system operation problems, the time interval is usually longer than one second, e.g., the time intervals of economic dispatch and unit commitment are both usually 1 hour, respectively. Therefore, it is necessary to know whether the proposed bilinear model is accurate for different time intervals. In this regard, the status of the air in the cavern is calculated using the procedure given in Figs. 4a, 4b, and 4c for different time intervals (1 second, 1 minute, 5 minutes, 10 minutes, 20 minutes, and 60 minutes) using the same initial status. The final temperature and pressure of the charging, discharging, and idle processes obtained by the bilinear model and the analytical model are plotted in Fig. 12. The error and relative error (values given in round brackets) between the results, in terms of final temperature and pressure, obtained from the two models are shown in Table IV, where the $2^{nd}$-$5^{th}$, the $6^{th}$-$9^{th}$, and the $10^{th}$-$13^{th}$ rows show the error/relative error in the charging, discharging, and idle processes, respectively. Again, the unit of temperature is set to Kelvin when calculating the relative error of temperature.

Figs. 12a-12d show that the accuracy of the bilinear model in the charging and discharging processes decreases as the time interval increases. This is because the charging and discharging parts of the model utilize the assumption that the air injected into or released from the cavern in each time step is much smaller than the total amount of air in the cavern (i.e., $\dot{m}_{in}\Delta t \ll m_s$ and $\dot{m}_{out}\Delta t \ll m_s$); this assumption introduces higher error when the time interval is larger. Figs. 12e-12f show that the accuracy of the bilinear model in the idle process does not change with the time interval; this is because the idle part of the model does not utilize the above assumption.

Tables IV-V show that the error and relative error of the temperature and pressure in both the charging and discharging processes are small when the time interval is less than or equal to 10 minutes. When the time interval is equal to 20 minutes, the errors (relative errors) of the pressure in the charging and discharging processes are 0.121 and 0.045 bar (0.17 and 0.09%), respectively, which are still relatively small. When the time interval is equal to 60 minutes, the error (relative error) in both the charging and discharging processes is relatively large. Table VI shows that the idle process of the model is quite accurate for all time intervals. Therefore, Fig. 12 and Tables IV-VI show that the accuracy of the bilinear cavern model is high, moderate, and relatively low when the time interval is between 1 second and 10 minutes, equal to 20 minutes, and equal to 60 minutes, respectively.

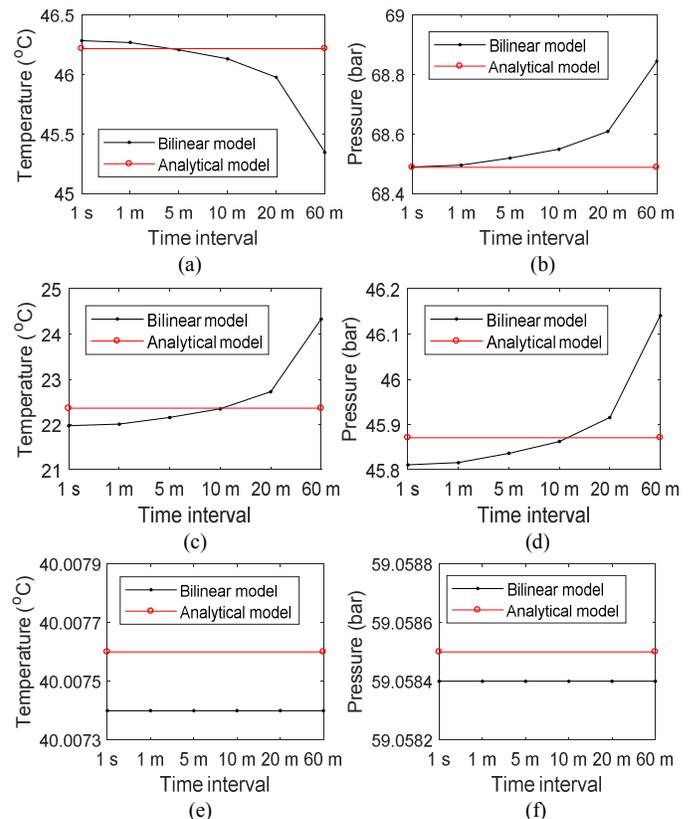

Fig. 12. Final temperature/pressure obtained by both the analytical model and the bilinear model in different processes using different time intervals: (a) final temperature in the charging process, (b) final pressure in the charging process, (c) final temperature in the discharging process, (d) final pressure in the discharging process, (e) final temperature in the idle process, and (f) final pressure in the idle process.

TABLE IV

ERROR (RELATIVE ERROR IN ROUND BRACKETS) BETWEEN THE SOLUTION OBTAINED BY THE BILINEAR MODEL AND THE ANALYTICAL MODEL IN THE CHARGING PROCESS USING DIFFERENT TIME INTERVALS

| Interval | 1 s | 1 min | 5 min | 10 min | 20 min | 60 min |
|---|---|---|---|---|---|---|
| Temperature (°C) | 0.066 | 0.051 | -0.010 | -0.087 | -0.241 | -0.872 |
| | (0.02%) | (0.02%) | (-3E-5) | (-0.03%) | (-0.08%) | (-0.27%) |
| Pressure (bar) | 0.002 | 0.007 | 0.031 | 0.061 | 0.121 | 0.356 |
| | (0.002%) | (0.01%) | (0.05%) | (0.09%) | (0.17%) | (0.52%) |

TABLE V

ERROR (RELATIVE ERROR IN ROUND BRACKETS) BETWEEN THE SOLUTION OBTAINED BY THE BILINEAR MODEL AND THE ANALYTICAL MODEL IN THE DISCHARGING PROCESS USING DIFFERENT TIME INTERVALS

| Interval | 1 s | 1 min | 5 min | 10 min | 20 min | 60 min |
|---|---|---|---|---|---|---|
| Temperature (°C) | -0.386 | -0.350 | -0.201 | -0.014 | 0.366 | 1.968 |
| | (-0.13%) | (-0.12%) | (-0.07%) | (-5E-5) | (0.12%) | (0.67%) |
| Pressure (bar) | -0.060 | -0.055 | -0.034 | -0.008 | 0.045 | 0.270 |
| | (-0.13%) | (-0.12%) | (-0.07%) | (-0.02%) | (0.09%) | (0.58%) |

TABLE VI

ERROR (RELATIVE ERROR IN ROUND BRACKETS) BETWEEN THE SOLUTION OBTAINED BY THE BILINEAR MODEL AND THE ANALYTICAL MODEL IN THE IDLE PROCESS USING DIFFERENT TIME INTERVALS

| Interval | 1 s | 1 min | 5 min | 10 min | 20 min | 60 min |
|---|---|---|---|---|---|---|
| Temperature (°C) | -2E-4 | -2E-4 | -2E-4 | -2E-4 | -2E-4 | -2E-4 |
| | (-1E-6) | (-1E-6) | (-1E-6) | (-1E-6) | (-1E-6) | (-1E-6) |
| Pressure (bar) | -1E-4 | -1E-4 | -1E-4 | -1E-4 | -1E-4 | -1E-4 |
| | (-2E-6) | (-2E-6) | (-2E-6) | (-2E-6) | (-2E-6) | (-2E-6) |

### E. Self-scheduling Problem of Compressed Air Energy Storage in Day-ahead Electricity Market

This subsection presents the results of solving the SSP of CAES using different cavern models. When the time interval is set to 60 minutes, the day-ahead SSP has 24 steps, which introduces a relatively small computational burden. However, the error is relatively large as shown in the previous subsection. When the time interval is set to 20 minutes, the error is relatively small and the day-ahead SSP has 72 steps. When the time interval is set to 10 minutes or smaller, the error decreases but the computational burden increases. Therefore, the time interval in the SSP is set to 20 minutes as a trade-off between the accuracy and computational burden.

#### 1) Comparison Among Models 1-3

Models 1 and 3, described in Sections III-A and III-C, respectively, are both solved by a popularly used MINLP solver, i.e., BARON [46], with the values of objective functions (i.e., the profits from the electricity market) given in the 3rd and 2nd columns of Table VII, respectively. Model 2 is solved by CPLEX with the objective function value given in the 4th column of Table VII. Models 1-3 are MIBLP, MILP, and MINLP models, respectively, as mentioned in Section III. These three models are solved using different numbers of time steps (i.e., $n_t$). The stopping criteria for both the BARON and CPLEX solvers are set to a maximum time of 8 hours and a gap of 0.1%; the solving process will terminate as soon as one of the two criteria is met.

Table VII shows that the objective function value obtained from Model 2 is higher than that from Models 1 and 3 for all three different numbers of steps. Note that a larger objective value represents a better solution as the goal of SSP is to maximize the profits. When the numbers of time steps are 10, 24, and 72, the objectives of Model 2 are 1.3, 7.4, and 3.6% higher than Model 3 and 1.3, 7.4, and 1.1% higher than Model 1, respectively.

The electricity market price, power output, and pressure in the 24-time-step case are plotted in Fig. 13. Note that the power output of the CAES in Fig. 13b equals the discharging power minus the charging power. The high price hours are 9-12 and 18-21. Fig. 13b indicates that the CAES generally discharges power at high price hours and charges power at low price hours to maximize the profit. The difference between the results of Models 1-3 lies in the power output in hours 1, 12, 18, and 20-22. A detailed comparison between Models 2 and 3 is given below. A similar analysis can be applied to Models 2 and 1 but is not provided.

The results of Model 2 discharge more power at hours 12 and 21 and charge more power at hour 20 compared to the results of Model 3, as detailed in Table VIII. In Table VIII, the last column provides the total profit in hours 12, 20, and 21 and clearly shows that Model 2 achieves higher profits than Model 3 and coincides with Table VII.

Linearizing the bilinear model into an MILP problem results in high accuracy as shown in [40], i.e., Model 2 is almost as accurate as Model 1. Therefore, the higher profit obtained from Model 2 indicates that a better solution has been obtained from Model 2 than Models 1 and 3 and the solutions obtained from Models 1 and 3 are not optimal. This is because MILP problems are easier to solve than MINLP or MIBLP problems. This shows the advantage of the proposed bilinear cavern model, which can be converted to an MILP problem and is therefore suitable for integration into optimization problems considering CAES.

TABLE VII

OBJECTIVE VALUES ($) OF THE SELF-SCHEDULING PROBLEM OF COMPRESSED AIR ENERGY STORAGE USING DIFFERENT CAVERN MODELS (NUMBERS IN BRACKETS ARE THE RELATIVE DIFFERENCE BETWEEN MODEL 2 AND MODEL $i$ ($i = 1,3$))

| Number of time steps ($n_t$) | Model 3 - MINLP [29] | Model 1 - MIBLP | Model 2 - MILP |
|---|---|---|---|
| 10 | 2.28E+5 (1.3%) | 2.28E+5 (1.3%) | 2.31E+5 |
| 24 | 2.51E+5 (7.4%) | 2.51E+5 (7.4%) | 2.71E+5 |
| 72 | 5.43E+5 (3.6%) | 5.57E+5 (1.1%) | 5.63E+5 |

TABLE VIII

THE PRICE AND THE SOLUTION OBTAINED BY MODELS 2 AND 3 IN HOURS 12, 20, AND 21 OF THE SELF-SCHEDULING PROBLEM OF COMPRESSED AIR ENERGY STORAGE

| | | Hour 12 | Hour 20 | Hour 21 | Total profit ($) |
|---|---|---|---|---|---|
| | Price ($) | 225.58 | 240.26 | 229.65 | -- |
| Model 3 | Power output (MW) | 154.49 | 40.0 | 40.0 | -- |
| | Profit ($) | 34386.4 | 9490.4 | 9066.0 | 52942.8 |
| Model 2 | Power output (MW) | 290.0 | -60.0 | 101.71 | -- |
| | Profit ($) | 64548.2 | -14595.6 | 23052.6 | 73005.2 |

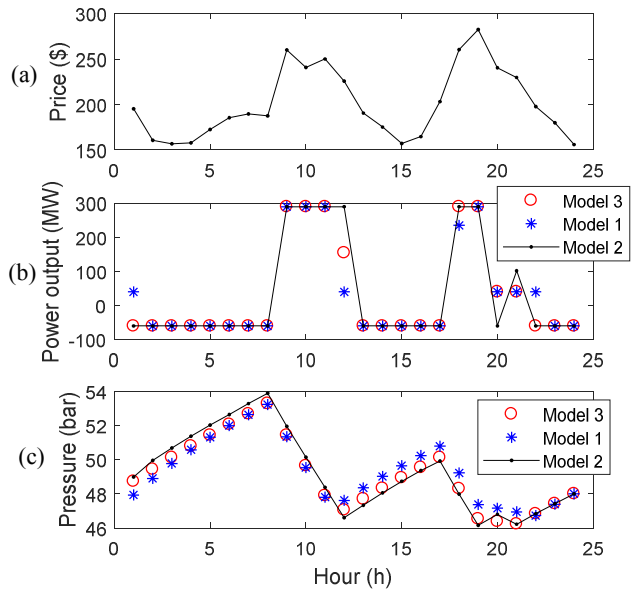

Fig. 13. The electricity market price (a) and the solution obtained by Models 1–3 in the 24-time-step case for the power output (b) and pressure (c).

### 2) Comparison Between the Bilinear and Constant-temperature Cavern Models

To show the advantage of the proposed bilinear cavern model over the constant-temperature model, both models are used to solve the SSP and comparative analysis is given as follows. The results, including the power output, temperature, and pressure, are given in Fig. 14. In Figs. 14b and 14c, the dots represent the temperature and pressure obtained from the SSP using the constant-temperature cavern model as detailed in Section III-C. Given the power output obtained from the SSP using the constant-temperature cavern model as input, the SSP using the bilinear cavern model is used to calculate the temperature and pressure as shown in the circles in Figs. 14b and 14c. That is, the circles given in Figs. 14b and 14c are the accurate values for the charging/discharging process given in Fig. 14a. Fig. 14b shows that the temperature during the charging/discharging process varies in a range of about 30 K, which is quite different from constant. In Fig. 14c, the pressure obtained from the constant-temperature cavern model varies from 46 to 56.43 bar (shown in dots); however, the pressure actually varies from 40.68 to 63.44 bar (shown in circles). At both hours 21 and 22, the pressure obtained from the constant-temperature cavern model is 46 bar; however, the pressure actually drops down to 40.68 and 40.79 bar, respectively, at these two time points. That is, the pressure obtained from the constant-temperature cavern model is quite inaccurate and the actual pressure drops far below the minimum bound of the normal operating range (i.e., 46 bar as shown in the dashed line in Fig. 14c). This inaccurate constant-temperature cavern model, however, has been adopted in a variety of power system application areas, e.g., [31], [32], [33], [34]. The proposed bilinear cavern model can replace the constant-temperature cavern model and is suitable for these power system applications and beyond, which shows the value of the proposed bilinear cavern model.

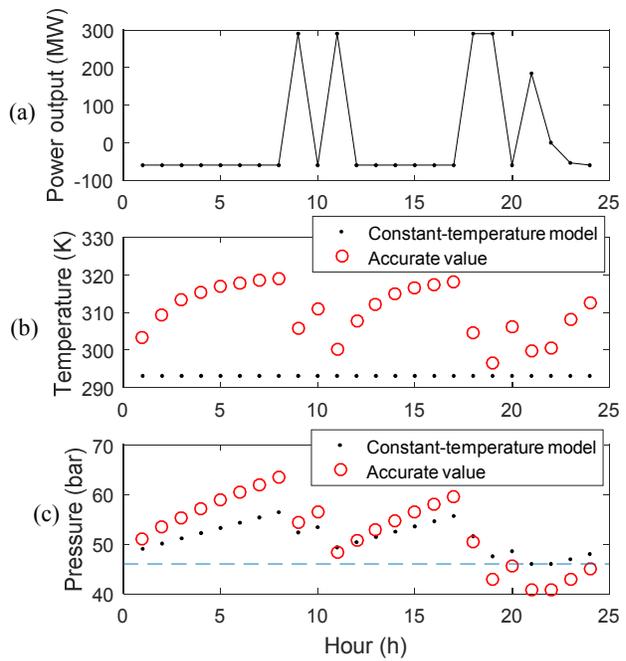

Fig. 14. The solution for the self-scheduling problem: (a) the power output obtained from the constant-temperature model, (b) the temperature obtained from the constant-temperature model and the corresponding accurate value, and (c) the pressure obtained from the constant-temperature model and the corresponding accurate value.

### F. Cavern Efficiency

The efficiency of the cavern is analyzed in this subsection. The internal energy is calculated from $mc_vT$, i.e., the product of mass of air $m$, a constant specific heat value $c_v$, and temperature $T$. Let the cavern efficiency be the ratio of the internal energy of all the air discharged out of the cavern in a period of time to the internal energy of all the air charged into the cavern in the same period of time. Note that the temperature of the air charged into the cavern is always equal to 50 °C because of the use of aftercoolers.

The initial temperatures of cavern air in the first day are set to 35, 40, and 45 °C, respectively. For each setting, Model 3 for the SSP of CAES in one day is solved; the internal energies of air injected into and released out of the cavern on this day are given in the 2nd and 4th columns of Table IX, respectively; and the heat transferred from the cavern wall to the cavern air on this day is given in the 3rd column of Table IX. Note that the positive (negative) values in the 3rd column of Table IX indicate that less (more) heat is transferred to the cavern wall than is received from the cavern wall.

The cavern efficiency for each setting is given in the last column of Table IX. Table IX shows that the cavern efficiency is higher than 1 when the initial temperature is 35 °C. This is because the temperature of the cavern wall (40 °C) is higher than 35 °C and thus heat is transferred to the cavern air from the cavern wall. That is, the cavern air gains energy from the cavern wall and therefore the cavern efficiency is higher than 1. But when the initial temperature is 40 or 45 °C, the cavern efficiency is lower than 1; this is because the temperature of cavern air is equal to or higher than that of the cavern wall and therefore heat is transferred to rather than received from the

    

cavern wall. That is, the cavern air loses energy to the cavern wall and therefore the cavern efficiency is lower than 1.

To determine the average cavern efficiency for a longer period of time, the efficiencies for 10 consecutive days for each setting are plotted in Fig. 15. For each of the $2^{nd}$ to $10^{th}$ days, the initial temperature of the cavern air in the next day is set to be the temperature of the cavern air at the end of the previous day and then the SSP of CAES in the day is solved to calculate the cavern efficiency for that day. The figure shows that the cavern efficiencies on the first day are quite different, but are close to one another for the $2^{nd}$-$10^{th}$ days. The average efficiency for all days is 0.9572 and the average efficiency for all days excluding the first day is 0.9579. Therefore, the average cavern efficiency is about 0.96.

TABLE IX
INTERNAL ENERGY OF AIR CHARGED IN AND DISCHARGED OUT, HEAT TRANSFERRED FROM CAVERN WALL, AND THE CAVERN EFFICIENCY

| Initial temperature of cavern air (°C) | Internal energy of air in (J) | Heat transferred from cavern wall (J) | Internal energy of air out (J) | Cavern efficiency |
|---|---|---|---|---|
| 35 | 3.45E+15 | 0.14E+15 | 3.59E+15 | 1.04 |
| 40 | 3.40E+15 | -0.17E+15 | 3.23E+15 | 0.95 |
| 45 | 3.42E+15 | -0.48E+15 | 2.94E+15 | 0.86 |

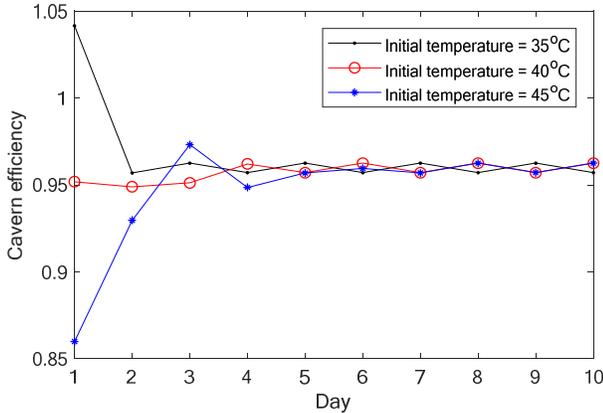

Fig. 15. Cavern efficiencies for 10 consecutive days using different initial temperatures for the first day.

## V. CONCLUSION

This paper proposed an accurate bilinear cavern model for compressed air energy storage based on the ideal gas law and the first law of thermodynamics. The accuracy of the proposed bilinear cavern model is verified via comparison with both an accurate analytical model in the literature and two sets of field-measured data.

Simulation results show that the mean absolute percentage error (mean absolute error) between the bilinear cavern model and the accurate analytical model for pressure (temperature) are no more than 0.82% and 0.36 bar (0.82% and 2.58 °C) for a variety of conditions. Note that the errors are usually less than the worst-case values given above. Simulation results also show that the pressure (temperature) results obtained from the bilinear cavern model can well match the pressure (temperature) from the two sets of field-measured data considered, i.e., datasets I and II, with a relative error of 0.47 and 1.01% (0.19 and 3.16%), respectively, in the normal operating mode of compressed air energy storage. The accuracy of the proposed bilinear model decreases as the time interval increases. For time intervals between 1 second and 10 minutes, equal to 20 minutes, and 60 minutes or longer, the bilinear cavern model has high, moderate, and relatively low accuracy, respectively. Simulation results also show that heat transfer has an obvious and measurable effect on the variation of temperature and pressure of the air in the cavern. Therefore, it is necessary to consider heat transfer in the cavern model. The constant-temperature cavern model is also shown to be inaccurate.

The self-scheduling problems of compressed air energy storage using different cavern models are solved. The simulation results show that the self-scheduling problem of compressed air energy storage using the proposed bilinear cavern model can be straightforwardly converted into a mixed-integer linear programming model that is easier to solve than the mixed-integer nonlinear programming. The self-scheduling problem of compressed air energy storage using the constant-temperature cavern model can result in infeasible solutions. However, by properly setting the time interval, the proposed bilinear cavern model is accurate and suitable for use in power system optimization problems and is superior to the existing highly nonlinear cavern model and the constant-temperature cavern model.

## APPENDIX

$a_1 = \frac{V_s}{c_v}(k-2)\frac{\Delta t^2}{2m_{av}}$,    $a_2 = \frac{R^k T_{in}^k}{V_s^k p_{in}^{k-1}}$,

$a_3 = \frac{R^{k-1} T_{in}^k}{V_s^{k-1} p_{in}^{k-1}}$    $a_4 = \frac{h_{eff} V_s t}{m_{av0} c_v}$,

$a_5 = \left(h_b \dot{m}_{in0}^{1.8} - h_b l_2 \dot{m}_{in0}\right)$,    $a_6 = \frac{V_s \Delta t}{c_v}$,

$a_7 = \frac{V_s \Delta t^2}{2m_{av} c_v}(k-1)$,    $a_8 = \frac{RT_{RW} \Delta t}{c_v}$,

$a_9 = \frac{R \Delta t^2}{2c_v}(k-1)$,    $a_{10} = \frac{R \Delta t}{c_v}$,

$a_{11} = \frac{\Delta t \Delta t}{c_v \rho_{av}}$,    $a_{12} = \frac{R T_{RW} \Delta t}{c_v}$,

$a_{13} = \frac{V_s}{2c_v} a_3 \Delta t^2 (m_{av0})^{k-3}$,

$c_0 = T_{in}^{\frac{k}{k-1}}/p_{in}$,    $c_1 = (T_{in2})^{\frac{k}{k-1}}/p_{in2}$,

$c_2 = (k-2)\Delta t - \frac{V_s}{c_v} h_b \Delta t\, l_1 - a_1 h_b l_2 - a_1 h_a$,

$c_3 = a_3 \Delta t(k-1)(m_{av0})^{k-2} - a_{13}(h_b l_2 + h_a)(k-2)$,

$c_4 = a_3 \Delta t (m_{av0})^{k-1} - \frac{V_s}{c_v} h_b T_{RW} \Delta t\, l_1 - a_3 \Delta t(k-1)(m_{av0})^{k-1} - a_{13}(h_b l_2 + h_a)(3-k) m_{av0}$,

$c_5 = \frac{V_s}{c_v}\left(h_b \Delta t l_1 \dot{m}_{in0} - h_b \Delta t \dot{m}_{in0}^{0.8} - h_a \Delta t\right) - a_1 h_b \dot{m}_{in0}^{1.8} + a_1 h_b l_2 \dot{m}_{in0}$,

$c_6 = -a_{13} a_5 (k-2)$,

$c_7 = \frac{V_s}{c_v} T_{RW} \left(\Delta t h_b \dot{m}_{in0}^{0.8} - \Delta t h_b\, l_1 \dot{m}_{in0} + h_a\, \Delta t\right) - a_{13} a_5 (3-k) m_{av0}$,

$c_8 = \left((k-1)\Delta t - a_{11} h_a - 0.5 a_{11}(k-2)(h_a + h_b\, l_2) - \Delta t V_s h_b l_1 / c_v - a_{11} h_b l_2\right)$,

$c_9 = (a_2 \Delta t k (m_{av0})^{k-1} + a_{12} h_b l_1)$,

$c_{10} = (a_2 \Delta t (1-k)(m_{av0})^k + a_{12} h_a \Delta t + a_{12} \Delta t h_b l_2)$,

$c_{11} = a_{12}\left(h_b \dot{m}_{in0}^{0.8} - h_b l_1 \dot{m}_{in0} + h_a\right)$,

$c_{12} = -\Big( \big(V_s h_b \dot{m}_{in0}{}^{0.8} + h_A V_s - V_s h_b l_1 \dot{m}_{in0}\big)\Delta t / c_v +$
$\qquad a_{11} h_b \big( \dot{m}_{in0}{}^{1.8} - l_2 \dot{m}_{in0} \big) + 0.5 a_{11}(k-2) h_b \big( \dot{m}_{in0}{}^{1.8} -$
$\qquad l_2 \dot{m}_{in0} \big) \Big),$

$c_{13} = a_{12}\Delta t h_b \big( \dot{m}_{in0}{}^{1.8} - l_2 \dot{m}_{in0} \big),$

$c_{14} = (a_7(h_a + h_b l_4) - (k-1)\Delta t - a_6 h_b l_3),$

$c_{15} = a_6 h_b l_3 T_{RW},$

$c_{16} = \big( a_7 \big( h_b \dot{m}_{out0}{}^{1.8} - h_b l_4 \, \dot{m}_{out0} \big) - a_6 \big( h_a + h_b \dot{m}_{out0}{}^{0.8} -$
$\qquad h_b l_3 \, \dot{m}_{out0} \big) \big),$

$c_{17} = a_6 T_{RW} \big( h_a + h_b \dot{m}_{out0}{}^{0.8} - h_b l_3 \, \dot{m}_{out0} \big),$

$c_{18} = -\big( k\Delta t + a_{10} \frac{V_s h_b}{R} l_3 \big),$

$c_{19} = -a_{10} h_a,$

$c_{20} = (a_{10}\Delta t h_a + a_9 h_a + (a_{10}\Delta t h_b + a_9 h_b) l_4),$

$c_{21} = -a_8 \Delta t h_b l_3,$

$c_{22} = a_8 h_b l_3,$

$c_{23} = -a_8 \Delta t (h_a + h_b \dot{m}_{out0}{}^{0.8} - h_b l_3 \dot{m}_{out0}),$

$c_{24} = a_8 (h_a + h_b \dot{m}_{out0}{}^{0.8} - h_b l_3 \dot{m}_{out0}),$

$c_{25} = -a_{10} \frac{V_s h_b}{R} \big( \dot{m}_{out0}{}^{0.8} - l_3 \dot{m}_{out0} \big),$

$c_{26} = (a_{10}\Delta t h_b + a_9 h_b)\big( \dot{m}_{out0}{}^{1.8} - l_4 \dot{m}_{out0} \big),$

$c_{27} = \frac{a_4 e^{-a_4}}{m_{avo}}, \qquad\qquad c_{28} = (e^{-a_4} - a_4 e^{-a_4}),$

$c_{29} = -T_{RW} \frac{a_4 e^{-a_4}}{m_{avo}}, \qquad c_{30} = -T_{RW}(e^{-a_4} - a_4 e^{-a_4} - 1),$

$c_{31} = -\frac{R T_{RW} a_4 e^{-a_4}}{m_{avo} V_s}, \qquad c_{32} = (1 - e^{-a_4} + a_4 e^{-a_4})\frac{R T_{RW}}{V_s}.$

## ACKNOWLEDGEMENTS

The work was supported in part by the Natural Sciences and Engineering Research Council (NSERC) of Canada and the Saskatchewan Power Corporation (SaskPower).